\documentclass[11pt]{article}

\def\ShowComment{True} 

\usepackage[utf8]{inputenc}
\usepackage[T1]{fontenc}

\usepackage{authblk}

\usepackage{geometry}
\usepackage[bottom]{footmisc}
\usepackage{todonotes}
\usepackage{xspace}

\usepackage{comment}

\usepackage{nicefrac}
\usepackage{algorithm}
\usepackage{algorithmicx}
\usepackage[noend]{algpseudocode}

\usepackage{amssymb}
\usepackage{amsmath}
\usepackage{amsthm}
\usepackage{dsfont}
\usepackage{footnote}

\usepackage{array}
\usepackage{verbatim}
\usepackage{multirow}
\usepackage{xcolor}
\usepackage{nameref}
\definecolor{ForestGreen}{rgb}{0.1333,0.5451,0.1333}
\definecolor{DarkRed}{rgb}{0.65,0,0}
\definecolor{Red}{rgb}{1,0,0}
\usepackage[linktocpage=true,
pagebackref=true,colorlinks,
linkcolor=DarkRed,citecolor=ForestGreen,
bookmarks,bookmarksopen,bookmarksnumbered]
{hyperref}
\usepackage{cleveref}

\usepackage{thm-restate}
\usepackage[tableposition=bottom]{caption}
\usepackage[all]{hypcap}
\usepackage{subcaption}
\usepackage{framed}
\usepackage[framemethod=tikz]{mdframed}
\usepackage[skins]{tcolorbox}

\makeatletter
\g@addto@macro{\maketitle}{\@thanks}
\makeatother

\newtheorem{thm}{Theorem}[section]
\newtheorem{cor}[thm]{Corollary}

\newtheorem{fact}[thm]{Fact}
\newtheorem{lem}[thm]{Lemma}
\newtheorem{definition}[thm]{Definition}

\newtheorem{obs}[thm]{Observation}
\newtheorem{claim}[thm]{Claim}
\newtheorem{question}[thm]{Question}

\newcommand{\E}{\mathbb{E}}%

\newcommand{\eps}{\epsilon}%
\newcommand{\p}{\textsc{P}}%
\newcommand{\poly}{\mathrm{poly}}
\newcommand{\polylog}{\mathrm{polylog}}%

\renewcommand{\algorithmiccomment}[1]{\bgroup\hfill$\rhd$~#1\egroup}

\newcounter{note}[section]

\newcommand{\Otil}{\tilde{O}}

\global\long\def\poly{\mathrm{poly}}
\global\long\def\polylog{\mathrm{polylog}}

\newenvironment{wrapper}[1]
{
	\begin{center}
		\begin{minipage}{\linewidth}
			\begin{mdframed}[hidealllines=true, backgroundcolor=gray!20, leftmargin=0cm,innerleftmargin=0.4cm,innerrightmargin=0.4cm,innertopmargin=0.4cm,innerbottommargin=0.4cm,roundcorner=10pt]
				#1}
			{\end{mdframed}
		\end{minipage}
	\end{center}
} 

\renewcommand{\paragraph}[1]{\medskip\noindent\textbf{#1}}

\ifdefined\ShowComment
\def\thatchaphol#1{\marginpar{$\leftarrow$\fbox{T}}\footnote{$\Rightarrow$~{\sf\textcolor{ForestGreen}{#1 --Thatchaphol}}}}

\def\sayan#1{\marginpar{$\leftarrow$\fbox{S}}\footnote{$\Rightarrow$~{\sf\textcolor{purple}{#1 --Sayan}}}}

\def\peter#1{\marginpar{$\leftarrow$\fbox{P}}\footnote{$\Rightarrow$~{\sf\textcolor{green}{#1 --Peter}}}}
\else
\def\thatchaphol#1{}
\def\sayan#1{}
\def\peter#1{}
\fi 

\usepackage{etoolbox}
\usepackage{tikz}
\usetikzlibrary{tikzmark}
\usetikzlibrary{calc}

\errorcontextlines\maxdimen

\newcommand{\ALGtikzmarkcolor}{black}
\newcommand{\ALGtikzmarkextraindent}{4pt}
\newcommand{\ALGtikzmarkverticaloffsetstart}{-.5ex}
\newcommand{\ALGtikzmarkverticaloffsetend}{-.5ex}
\makeatletter
\newcounter{ALG@tikzmark@tempcnta}

\newcommand\ALG@tikzmark@start{%
	\global\let\ALG@tikzmark@last\ALG@tikzmark@starttext%
	\expandafter\edef\csname ALG@tikzmark@\theALG@nested\endcsname{\theALG@tikzmark@tempcnta}%
	\tikzmark{ALG@tikzmark@start@\csname ALG@tikzmark@\theALG@nested\endcsname}%
	\addtocounter{ALG@tikzmark@tempcnta}{1}%
}

\def\ALG@tikzmark@starttext{start}
\newcommand\ALG@tikzmark@end{%
	\ifx\ALG@tikzmark@last\ALG@tikzmark@starttext
	\else
	\tikzmark{ALG@tikzmark@end@\csname ALG@tikzmark@\theALG@nested\endcsname}%
	\tikz[overlay,remember picture] \draw[\ALGtikzmarkcolor] let \p{S}=($(pic cs:ALG@tikzmark@start@\csname ALG@tikzmark@\theALG@nested\endcsname)+(\ALGtikzmarkextraindent,\ALGtikzmarkverticaloffsetstart)$), \p{E}=($(pic cs:ALG@tikzmark@end@\csname ALG@tikzmark@\theALG@nested\endcsname)+(\ALGtikzmarkextraindent,\ALGtikzmarkverticaloffsetend)$) in (\x{S},\y{S})--(\x{S},\y{E});%
	\fi
	\gdef\ALG@tikzmark@last{end}%
}

\apptocmd{\ALG@beginblock}{\ALG@tikzmark@start}{}{\errmessage{failed to patch}}
\pretocmd{\ALG@endblock}{\ALG@tikzmark@end}{}{\errmessage{failed to patch}}
\makeatother

\title{Sublinear Algorithms for $(1.5+\epsilon)$-Approximate  Matching}
\author[1]{Sayan Bhattacharya \footnote{Supported by Engineering and Physical Sciences Research Council, UK (EPSRC) Grant EP/S03353X/1.}}
\author[1,2]{Peter Kiss}
\author[3]{Thatchaphol Saranurak}
\affil[1]{University of Warwick}
\affil[2]{Max-Planck-Institut für Informatik}
\affil[3]{University of Michigan}

\date{}

\begin{document}

\maketitle

\pagenumbering{gobble}
\begin{abstract}
We study sublinear time algorithms for estimating the size of maximum matching. After a long line of research, the problem was finally settled by Behnezhad~{[}FOCS'22{]}, in the regime where one is willing to pay an approximation factor of $2$. Very recently, Behnezhad et al.~{[}SODA'23{]} improved the approximation factor to $(2-\frac{1}{2^{O(1/\gamma)}})$ using $n^{1+\gamma}$ time. This improvement over the factor $2$ is, however, minuscule and they asked if even $1.99$-approximation is possible in $n^{2-\Omega(1)}$ time.

We give a strong affirmative answer to this open problem by showing $(1.5+\epsilon)$-approximation algorithms that run in $n^{2-\Theta(\epsilon^{2})}$ time. Our approach is conceptually simple and diverges from all previous sublinear-time matching algorithms: we show a sublinear time algorithm for computing a variant of the \emph{edge-degree constrained subgraph} (EDCS), a concept that has previously been exploited in dynamic  {[}Bernstein Stein ICALP'15, SODA'16{]}, distributed  {[}Assadi et al. SODA'19{]} and streaming  {[}Bernstein ICALP'20{]} settings, but never before in the sublinear setting.

\medskip

\textbf{Independent work:} Behnezhad, Roghani and Rubinstein [BRR'23] independently showed sublinear algorithms similar to our Theorem 1.2 in both adjacency list and matrix models. Furthermore, in [BRR'23], they show additional results on strictly better-than-1.5 approximate matching algorithms in both upper and lower bound sides.

\end{abstract}

\pagebreak{}

\pagenumbering{arabic}
\section{Introduction}
\label{sec:intro}

Computing a maximum matching in a graph is one of the most fundamental problems in combinatorial optimization.   We consider this problem in {\em a sublinear setting},  as defined below.

\medskip
\noindent {\bf The model.} The input graph $G = (V, E)$ has $n$ nodes, maximum degree $\Delta$ and average degree $d$. The node-set $V$ is known to the algorithm. In contrast, the algorithm can access the edge-set $E$ by making either only {\em adjacency-matrix} queries, or  only {\em adjacency-list} queries. An adjacency-matrix query takes an unordered pair of vertices $\{u,v\}\in{V \choose 2}$, and returns whether $\{u,v\}\in E$. An adjacency-list query, on the other hand, takes an ordered pair $(v,i)$, where $v\in V$ and $i\in[n]$, and returns the $i^{th}$ edge incident on $v$ in $G$ if $i \le \deg_v(E)$, or returns $\perp$ if $i > \deg_{v}(E)$.   We say that a matching $M \subseteq E$ is $(\alpha, \beta)$-approximate iff $\mu(G) \leq \alpha \cdot |M| + \beta$, where $\mu(G)$ is the size of maximum matching in $G$. When $\beta = 0$,  we refer to an $(\alpha, 0)$-approximate matching as simply an $\alpha$-approximate matching.
 In this model, the goal  is to design  algorithms  that runs in {\em sublinear}  time (i.e., without even reading the entire input), assuming that it takes $O(1)$ time to answer each query.

\medskip
Before proceeding any further, we highlight a known fact (see Appendix~\ref{appendix:fact} for  more details). 

\begin{wrapper}
\begin{fact}(Informal)\label{fact:intro:main}
No sublinear algorithm  returns the edges of a good approximate  matching.
\end{fact}
\end{wrapper}

To see an intuitive justification behind Fact~\ref{fact:intro:main}, consider a $(n/2) \times (n/2)$ bipartite graph, whose edge-set is a perfect matching given by a uniformly random permutation of $[n/2]$. In this graph,  no algorithm can return a $(O(1), \epsilon n)$-approximate maximum matching  with $o(n^2)$ adjacency-matrix queries. A similar instance gives the desired lower bound under adjacency-list queries.

Accordingly, for over a decade, a long and impressive line of work has been devoted to the study of the {\em value version} of this problem, where the sublinear algorithm is asked to return only an approximation to the {\em value} of $\mu(G)$. The following remains a central open question in this topic.\footnote{Throughout this paper, the notation $\tilde{O}(.)$ will be used to hide $\poly\log n$ factors, whereas the notation $O_{\epsilon}(.)$ will hide any function that solely depends on $\epsilon$.
}
\begin{wrapper}
\begin{question}
\label{q:main}
      Under adjacency-matrix queries, 
      can we design a $(1, \epsilon n)$-approximation algorithm in $\tilde{O}_{\epsilon}\left(n^{2-\Theta(1)}\right)$ time?
     Similarly, under adjacency-list queries,  can we design a $(1+\epsilon)$-approximation algorithm in $\tilde{O}_{\epsilon}\left(n \Delta^{1-\Theta(1)}\right)$ time?
    \end{question}
\end{wrapper}
Note that we need $\Omega(n^2)$  adjacency-matrix queries to distinguish between an empty graph and a graph containing one edge. This explains why Question~\ref{q:main} asks for an $(1, \epsilon n)$-approximation under adjacency matrix queries, but a multiplicative $(1+\epsilon)$-approximation under adjacency-list queries.

\medskip
\noindent {\bf Previous Work.} We summarize the relevant literature in  \Cref{tab:summary}. Early research on this topic focused on obtaining $O(1)$ time algorithms when $\Delta = O(1)$. For instance, a notable paper by Yoshida~et~al.~\cite{YoshidaYI12} designed a $(1,\epsilon n)$-approximation in $\Delta^{O(1/\epsilon^{2})}$ time under adjacency-list queries. On general graphs, however, all these early algorithms \cite{parnas2007approximating,nguyen2008constant,YoshidaYI12,onak2012near} can take $\Omega(n^{2})$ time. This drawback was addressed   by Kapralov et al.~\cite{kapralov2020space} and Chen et al.~\cite{chen2020sublinear} (based on \cite{onak2012near}), who obtained the first nontrivial bounds for this problem on general graphs. Continuing this line of work, in a recent influential paper Behnezhad \cite{Behnezhad21} finally designed sublinear algorithms for matching with near-optimal running times in general graphs, in the regime where the approximation guarantee is arbitrarily close to $2$. The key idea behind the result of Behnezhad \cite{Behnezhad21} is to implement the randomized greedy algorithm, which returns a maximal matching and has an approximation guarantee of $2$, in the sublinear setting.

At this point, a natural question to ask is whether one can beat the performance of randomized greedy matching  in the sublinear model. This  was answered very recently by Behnezhad et al.~\cite{behnezhad2023beating}, who showed that for any constant $\gamma>0$, the approximation guarantee of~\cite{Behnezhad21} can be improved to $2-\frac{1}{2^{O(1/\gamma)}}$, by paying an additional factor of $\Delta^{\gamma}$ in running time. This improvement over the approximation guarantee of $2$ is, however, minuscule, and it was explicitly posted as an open problem in \cite{behnezhad2023beating} whether even $1.99$-approximation is possible in $n^{2-\Omega(1)}$ time.

\begin{table}
\footnotesize{

\begin{tabular}{|>{\raggedright}p{0.1\textwidth}|>{\centering}p{0.15\textwidth}|>{\centering}p{0.1\textwidth}|>{\centering}p{0.1\textwidth}|>{\centering}p{0.1\textwidth}|>{\centering}p{0.15\textwidth}|>{\centering}p{0.1\textwidth}|}
\hline 
Model & \multicolumn{2}{c|}{Adjacency List } & \multicolumn{2}{c|}{Adjacency List } & \multicolumn{2}{c|}{Adjacency Matrix}\tabularnewline
\hline 
Guarantee & Approx & Time & Approx & Time & Approx & Time\tabularnewline
\hline 
\hline 
\cite{parnas2007approximating} & $(2,\epsilon n)$ & $\Delta^{O(\log(\Delta/\epsilon)}$ &  &  &  & \tabularnewline
\hline 
\multirow{2}{0.1\textwidth}{\cite{nguyen2008constant}} & $(2,\epsilon n)$ & $2^{O(\Delta)}/\epsilon^{2}$ &  &  &  & \tabularnewline
\cline{2-7} 
 & $(1,\epsilon n)$ & $2^{\Delta^{O(1/\epsilon)}}$ &  &  &  & \tabularnewline
\hline 
\multirow{2}{0.1\textwidth}{\cite{YoshidaYI12}} & $(2,\epsilon n)$ & $\Delta^{4}/\epsilon^{2}$ &  &  &  & \tabularnewline
\cline{2-7} 
 & $(1,\epsilon n)$ & $\Delta^{O(1/\epsilon^{2})}$ &  &  &  & \tabularnewline
\hline 
\cite{onak2012near,chen2020sublinear} & $(2,\epsilon n)$ & $(d+1)\Delta/\epsilon^{2}$  &  &  & $(2,\epsilon n)$ & $n\sqrt{n}/\epsilon^{2}$ \tabularnewline
\hline 
\cite{kapralov2020space} & $(O(1),\epsilon n)$ & $\Delta/\epsilon^{2}$ & $O(1)$ & $n+\Delta^{2}/d$ &  & \tabularnewline
\hline 
\cite{Behnezhad21} & $(2,\epsilon n)$ & $(d+1)/\epsilon^{2}$ & $2+\epsilon$ & $n+\Delta/\epsilon^{2}$ & $(2,\epsilon n)$ & $n/\epsilon^{3}$\tabularnewline
\hline 
\cite{behnezhad2023beating} & $(2-\frac{1}{2^{O(1/\gamma)}},o(n))$ & $(d+1)\Delta^{\gamma}$ & $2-\frac{1}{2^{O(1/\gamma)}}$ & $n+\Delta^{1+\gamma}$ & $(2-\frac{1}{2^{O(1/\gamma)}},o(n))$ & $n^{1+\gamma}$\tabularnewline
\hline 
\textbf{Ours} & $(1.5,\epsilon n)$ & $nd^{1-\Theta(\epsilon^{2})}$ & $1.5+\epsilon$ & $n\Delta^{1-\Theta(\epsilon^{2})}$ & $(1.5,\epsilon n)$ & $n^{2-\Theta(\epsilon^{2})}$\tabularnewline
\hline 
\end{tabular}

}

\caption{Summary of sublinear  algorithms for maximum matching. We omit $\protect\polylog(n/\epsilon)$ factors. 
\label{tab:summary}}
\end{table}

\paragraph{Our Results.}
We give a strong affirmative answer to this open problem by showing the first sublinear matching algorithms with approximation guarantee arbitrarily close to $1.5$.
\begin{wrapper}
\begin{thm}
\label{thm:main}For any $\epsilon>0$, there are algorithms that, given a graph $G$, w.h.p. estimates the size of maximum matching up to
\begin{enumerate}
\item $(1.5+\epsilon)$-approximation in $\Otil_{\epsilon}\left(n\Delta^{1-\Theta(\epsilon^{2})}\right)$ time using adjacency-list queries,
\item $(1.5,\epsilon n)$-approximation in $\Otil_{\epsilon}\left(nd^{1-\Theta(\epsilon^{2})}\right)$ time using adjacency-list queries, and
\item $(1.5,\epsilon n)$-approximation in $\Otil_{\epsilon}\left(n^{2-\Theta(\epsilon^{2})}\right)$ time using using adjacency-matrix queries.
\end{enumerate}
\end{thm}
\end{wrapper}

 \Cref{thm:main} makes  significant progress towards the ultimate goal of resolving Question~\ref{q:main}. An affirmative resolution to this question would have impact beyond the area of sublinear algorithms. For example, a $(1,\epsilon n)$-approximation  algorithm in $n^{2-\Theta(1)}$ time, under adjacency-matrix queries, would immediately imply the first fully dynamic algorithm that can $(1+\epsilon)$-approximate the size of maximum matching using $n^{1-\Theta(1)}$ update time, which is one of the biggest open problems in the area of dynamic graph algorithms~\cite{Behnezhad23,bhattacharya2023dynamic}. See, e.g.~\cite{bhattacharya2023dynamic}, for the reduction.

\paragraph{Our Techniques.}
Our technique behind \Cref{thm:main} diverges from all previous sublinear matching algorithms. In brief, we show how to exploit the concept of an  \emph{edge-degree constrained subgraph} (EDCS) in the sublinear model. This object has been successfully used in the past in dynamic  \cite{BernsteinS15,BernsteinS16,GrandoniSSU22,Kiss22}, distributed  \cite{AssadiBBMS19} and  streaming settings \cite{Bernstein20,AssadiB21}, but never before in a sublinear setting. 

More specifically, our approach is inspired by the streaming algorithm of~\cite{Bernstein20}. We sample a sublinear number of edges from the input graph $G$, and using a greedy heuristic find a subset $H$ of these sampled edges which satisfy: $\deg_{u}(H) + \deg_v(H) \leq \beta$ for all $(u, v) \in H$, where $\deg_x(H)$ is the degree of node $x$ in $H$ and $\beta := 1/\Theta(\epsilon^3)$ can be thought of as being a large constant. We refer to $H$ as an {\em edge degree bounded subgraph} (EDBS) of $G$. Let $U := \{ (u, v) \in E \setminus H : \deg_u(H) + \deg_v(H) \leq (1-\epsilon) \beta \}$ denote the set of {\em underfull} edges. It is known~\cite{Bernstein20} that $U \cup H$ serves as a good {\em sparsifier} for maximum matching in the input graph, in the sense that $\mu(G) \leq (3/2+\epsilon) \cdot \mu(H \cup U)$.

The key observation driving our  algorithm is this: If the number of edges we sample from $G$ to construct $H$ is sufficiently large (while still being sublinear), then the average degree in $U$ is small. Since the EDBS $H$ has maximum degree $\leq \beta$ by definition, it follows that the subgraph $H \cup U$ also has small maximum degree. Thus, we can run the algorithm from~\cite{YoshidaYI12} on $H \cup U$, which works well on bounded degree graphs, to obtain a $(1+\epsilon)$-approximation to $\mu(H \cup U)$, which in turn gives a $(3/2+\epsilon)$-approximation to $\mu(G)$. One issue here is that the algorithm from~\cite{YoshidaYI12} requires adjacency-list query access to its input, which we cannot provide because we do not explicitly store the edges of $U$. We overcome this challenge by noting that, intuitively, we have adjacency-matrix query access to $H \cup U$ (because we do in fact explicitly store the set $H$). Thus, we can simulate an adjacency-list query $(v, i)$ in $H \cup U$, by first querying  $G$ to collect all the edges in $G$ that are incident on $v$, and then deciding which of these edges belong to $U$. This implies that we incur a blow up in the running time of \cite{YoshidaYI12} while implementing it on $H \cup U$. Nevertheless, by carefully choosing the relevant parameters, we still manage to ensure that the overall running time of our algorithm remains sublinear. The basic template behind our algorithms is  conceptually simple, and is described in Section~\ref{sec:schematic}. Section~\ref{sec:implementation} shows how to implement this template in the sublinear setting.

\paragraph{Remarks.}
We now point out an intriguing implication of our result. We can show that the subsets $H, U$, as described above, satisfy: either $\mu(H) \geq \Theta(\epsilon) \cdot \mu(G)$ or $\mu(H \cup U) \geq (1-\Theta(\epsilon)) \cdot \mu(G)$. In the former case, we can actually return the edges of a $1/(\Theta(\epsilon))$-approximate matching, since we explicitly store the edge-set $H$. In the latter case, we can return a $(1+\epsilon)$-approximation to the value of $\mu(G)$. This implies that on any given graph $G$, our algorithms  either refutes Fact~\ref{fact:intro:main}, or resolves Question~\ref{q:main} in the affirmative. This is summarized in Theorem~\ref{thm:intro:dichotomy}. See Section~\ref{sec:dichotomy} for details.

\begin{wrapper}
\begin{thm}
\label{thm:intro:dichotomy} There is an algorithm for each of the following tasks, on  input graph $G = (V, E)$.
\begin{itemize}
    \item (i) Return either a matching $M \subseteq E$ of size $|M| = \Omega(\epsilon \cdot \mu(G))$, or  a $(1, \epsilon n)$-approximation to the value of $\mu(G)$, in  $\tilde{O}_{\epsilon}\left( n^{2-\Theta(\epsilon^2)}\right)$ time under  adjacency-matrix queries, whp.
    \item (ii) Return either a matching $M \subseteq E$ of size $|M| = \Omega(\epsilon \cdot \mu(G))$, or a $(1+ \epsilon )$-approximation to the value of $\mu(G)$, in  $\tilde{O}_{\epsilon}\left( n \cdot \Delta^{1-\Theta(\epsilon^2)}\right)$ time under adjacency-list queries, whp.
    \item (iii) Return either a matching $M \subseteq E$ of size $|M| = \Omega(\epsilon \cdot \mu(G))$, or  a $(1, \epsilon n)$-approximation to the value of $\mu(G)$, in  $\tilde{O}_{\epsilon}\left(n \cdot \left(1+d^{1-\Theta(\epsilon^2)}\right)\right)$ time under adjacency-list queries, whp.
\end{itemize}
\end{thm}
\end{wrapper}

Also, note that we are using the algorithm from~\cite{YoshidaYI12} (which requires adjacency-list query access) on  $H \cup U$ (which we can access via only adjacency-matrix queries). If there exists a fast $(1,\epsilon n)$-approximate sublinear algorithm under adjacency-matrix queries, then we can use it directly to estimate the value of $\mu(H \cup U)$. This will lead to an improved running time for our overall algorithm, as captured in the corollary below. In particular, an adjacency-matrix query based $(1, \epsilon n)$-approximation algorithm in $\tilde{O}_{\epsilon}(n \Delta)$ time, which is \textit{not} even sublinear in $|E|$,  will imply a $(1.5, \epsilon n)$-approximation algorithm, also under adjacency-matrix queries, in $\tilde{O}_{\epsilon}(n^{1.5})$ time.

\begin{wrapper}
\begin{cor}
\label{cor:intro:main:adj-matrix}
Suppose that there is an algorithm  which, given an input graph $G$, whp returns a $(1, \epsilon n)$-approximation to the value of $\mu(G)$ in  $\tilde{O}_{\epsilon}(n \Delta^{q})$ time under adjacency-matrix queries, where $q \geq 0$. Then there exists another adjacency-matrix query based algorithm which returns a $(1.5, \epsilon n)$-approximation to $\mu(G) $ in  $\tilde{O}_{\epsilon}\left(n^{2-1/(1+q)}\right)$ time.
\end{cor}
\end{wrapper}

Finally, as an immediate corollary of our result, we obtain a sublinear algorithm for graphic TSP with an improved approximation ratio from $27/14 \approx 1.929$ \cite{chen2020sublinear,behnezhad2022time}  to $40/21 \approx 1.904$.  See Appendix~\ref{appendix:tsp} for further discussions.

\begin{wrapper}
\begin{thm}
\label{thm:intro:tsp}
There is an $\tilde{O}(n^{2-\Theta(\epsilon^2)})$ running time randomized algorithm which estimates the cost of graphic TSP within a factor of $40/21+\epsilon$.
\end{thm}
\end{wrapper}

\section{Preliminaries}

\label{sec:prelims}

\noindent {\bf Key Notations:} Throughout this paper, we denote the input graph by $G = (V, E)$, and let $m, n$, $\Delta$ and $d$ respectively denote the number of edges, the number of nodes, the maximum degree of any node, and the average degree of a node in $G$.   For a subset of edges $E' \subseteq E$, we often abuse the notation and refer to $E'$ as a subgraph $G' = (V, E')$ of $G$.   Next, consider any edge $e = (u, v) \in {V \choose 2}$ that is not necessarily part of $G$. The {\em degree} of this edge  in $G$ is defined as $\deg_e(E) := \deg_u(E) + \deg_v(E)$, where $\deg_x(E)$ denotes the degree of a node $x \in V$ in $G = (V, E)$.  Furthermore, we define
\begin{equation}
    \label{eq:parameters}
    \beta := \frac{1}{\Theta(\epsilon^3)}, \text{ where } \epsilon \in (0, 1) \text{ is a small constant.}
\end{equation}

\medskip
\noindent {\bf EDBS:}
 We now recall the notion of an {\em edge degree constrained subgraph} (EDCS) of the input graph, which in recent years has found a wide variety of applications across different computational models~\cite{AssadiB21,AssadiB19,AssadiBBMS19,Bernstein20,BernsteinS15,BernsteinS16,GrandoniSSU22,Kiss22}.  Specifically, an EDCS $H \subseteq E$ is a subgraph of $G = (V, E)$ that satisfy two conditions: (i) Each edge $e \in H$ inside the EDCS has degree $\deg_e(H) \leq \beta$, and (ii) each edge $e \in E \setminus H$ outside of the EDCS has degree $\deg_e(H) \geq (1-\epsilon)\beta$.  It is known that $\mu(H) \leq \mu(G) \leq (3/2+\epsilon) \cdot \mu(G)$ for any EDCS $H$ of $G$. For reasons that will become apparent as we describe our algorithm, in this paper we drop the second condition from the definition of an EDCS. Instead, we say that $H \subseteq E$ is an {\em edge degree bounded subgraph} (EDBS) of $G$ iff  $\deg_e(H) \leq \beta$ for all $e \in H$. Theorem~\ref{thm:EDCS} implies that any EDBS $H$ of $G$, along with all the edges not in $H$ which violate the second condition mentioned above, still preserves a $(3/2+\epsilon)$-approximately maximum. This relaxation of the notion of an EDCS was considered before in~\cite{Bernstein20}, albeit in the different context of  maximum matching in the semi-streaming model.

\begin{definition}
\label{def:EDBS}  Consider any subset of edges $H \subseteq E$ in an input graph $G = (V, E)$. We say that an edge $e \in E$ is {\em underfull} w.r.t.~$H$ iff $\deg_e(H) < (1-\epsilon) \beta$, and {\em overfull} iff $\deg_e(H) > \beta$. Furthermore, we say that $H$ is an {\em EDBS} of $G$ iff there is no edge $e \in H$ that is overfull w.r.t.~$H$. 
\end{definition}

\begin{thm}[\cite{Bernstein20}]
\label{thm:EDCS} Consider any graph $G = (V, E)$, and an EDBS $H$ of $G$. Let $U$ denote the collection of edges in $E \setminus H$ that are underfull w.r.t.~$H$. Then $\mu(G) \leq (3/2+\epsilon) \cdot \mu(H \cup U)$.
\end{thm}

Next, consider a process whereby we start with an empty subset of edges  $H = \emptyset$ in  $G$, and keep making the following two types of changes to $H$: (1)  if we find an edge $e \in E \setminus H$ that is underfull w.r.t.~$H$, then we include it in $H$, and (2) If we find an edge $e \in H$ that is overfull w.r.t.~$H$, then we remove it from $H$. Theorem~\ref{thm:local-moves} implies that this  can continue for at most $O(\beta^2 \cdot \mu(G))$ iterations.

\begin{definition}
\label{def:local-move}
The operation {\sc Insert}$(H, e, G)$ takes as input a graph $G = (V, E)$, a subset of edges $H \subseteq E$ and an edge $e \in E \setminus H$ that is underfull w.r.t.~$H$, and it outputs the subset $H \cup \{ e \}$. In contrast, the operation {\sc Delete}$(H, e, G)$ takes as input a graph $G = (V, E)$, a subset of edges $H \subseteq E$ and an edge $e \in H$ that is overfull w.r.t.~$H$, and it outputs the subset $H \setminus \{e\}$.
\end{definition}

\begin{thm}
\label{thm:local-moves}
Consider a process where we start with an empty subset of edges $H = \emptyset$ in $G = (V, E)$. Subsequently, the set $H$ changes via a sequence of {\sc Insert} and {\sc Delete} operations. To be more specific, in each iteration of this process, either we find an edge $e \in E \setminus H$ that is underfull w.r.t.~$H$ and set $H \leftarrow \text{{\sc Insert}}(H, e, G)$, or we find an edge $e \in H$ that is overfull w.r.t.~$H$ and set $H \leftarrow \text{{\sc Delete}}(H, e, G)$.  Then this process can run for at most $O(\beta^2 \cdot \mu(G))$ iterations.
\end{thm}

Theorem~\ref{thm:local-moves} follows  from the standard potential function based argument for EDCS~\cite{AssadiBBMS19,AssadiB19,BernsteinS16}. For completeness, we provide a proof of this theorem in Appendix~\ref{appendix:proof:theorem:edcs}.

\medskip

The {\em query complexity} of an algorithm on a given input is the number of queries it needs to make, whereas its running time is the total time taken to return an answer to the given input, assuming each query is answered in $O(1)$ time. Note that the runtime of an algorithm is always lower bounded by its query complexity. With adjacency-matrix (resp.~adjacency-list) queries, it is trivial to design an algorithm with query complexity $O(n^2)$ (resp.~$O(m)$) for any graph problem. Our goal is to design an algorithm that returns an approximation to the value of $\mu(G)$, whose query complexity (and runtime) is much better than this trivial bound. We will use two well-known results from the sublinear matching literature, which are stated below.

\begin{thm}[\cite{YoshidaYI12}]
\label{thm:Yoshida}  There exists an algorithm that returns a $(1+\epsilon)$-approximation to the value of $\mu(G)$, where $G = (V, E)$ is the input graph, using $\tilde{O}_{\epsilon}\left(\Delta^{\Theta(\epsilon^{-2})}\right)$ adjacency-list queries whp. The algorithm also runs in  $\tilde{O}_{\epsilon}\left(\Delta^{\Theta(\epsilon^{-2})}\right)$ time whp.\footnote{Throughout this paper, we use the abbreviation ``whp'' to refer to the phrase ``with high probability''.} 
\end{thm}

\begin{thm}[\cite{Behnezhad21}]
\label{thm:Soheil}  There exists an algorithm that returns a $(2+\epsilon)$-approximation to the value of $\mu(G)$, where $G = (V, E)$ is the input graph, using $\tilde{O}_{\epsilon}(n)$ adjacency-list queries whp. The algorithm also runs in  $\tilde{O}_{\epsilon}(n)$  time whp.
\end{thm}

\section{A Schematic Algorithm}
\label{sec:schematic}

We now  describe our algorithmic template (see Figure~\ref{alg:sample}), which takes as input, along with the graph $G = (V, E)$, the parameters $\mu^*, m^*, \Delta^*$ and $\gamma$. The reader should think of the first three parameters in this list as being proxies for $\mu(G), m$ and $\Delta$ (or sometimes, $d$ instead of $\Delta$).\footnote{Recall that $d$ and $\Delta$ respectively denotes the average degree and maximum degree in $G$.} We  choose different values for these parameters while implementing the  algorithm in  different settings -- under adjacency-list and adjacency-matrix queries. However, in all these settings, we  have:
\begin{equation}
    \label{eq:parameters}
    \frac{\mu(G)}{(2+\epsilon)} \leq \mu^* \leq n, \ d \leq \Delta^* \leq n, \ m^* \geq m, \text{ and } 0 < \gamma < 1.
\end{equation}
 The algorithm proceeds in {\em rounds}. In each round, it samples $\Theta\left(\frac{m^* \log n}{\mu^* (\Delta^*)^{\gamma}}\right)$ edges independently and uniformly at random from the set $E$ (with repetitions). For each sampled edge $e \in E$, the algorithm checks if $e \in E \setminus H$ and is underfull w.r.t.~$H$. If yes, then it inserts $e$ into $H$, which in turn might lead to some existing edges in $H$ becoming overfull. Before dealing with the next sample, the algorithm ensures that  $H$ remains an EDBS of $G$, by repeatedlty removing overfull edges from $H$ if necessary. Finally, if  the set $H$ does not change  throughout the entire duration of a given round, then the algorithm no longer proceeds to the next round. Instead, at this point it identifies the collection $U \subseteq E \setminus H$ of edges that are underfull w.r.t.~$H$ (but does not explicitly calculate $U$), identifies the nodes  $V_{small} \subseteq V$ that have ``small''  degrees in $U$, and then returns the size of the maximum matching in the subgraph of $H \cup U$ induced by $V_{small}$. In the lemmas below, we  derive a few key properties of this schematic algorithm.

\begin{figure*}[htbp]
\centerline{\framebox{
    \begin{minipage}{5.5in}
        \begin{tabbing}
            01.  \  \=   $H \leftarrow \emptyset$. \\          
            02.  \> {\sc Repeat}:  \qquad \qquad  \qquad \qquad \qquad \qquad // (Start of a new round)\\
            03.  \> \ \ \ \ \ \ \ \= $\text{{\sc Status}} \leftarrow \text{false}$. \\
            04. \> \> {\sc For} $i = 1$ to $(100  m^*  \log n)/(\mu^* (\Delta^*)^{\gamma})$ \\
            05. \> \> \ \ \ \ \ \ \= Sample an edge $e \in E$ u.a.r. \\
            06. \> \> \> {\sc If} $e \in E \setminus H$ and is underfull w.r.t.~$H$, {\sc Then} \\
            07. \> \> \> \ \ \ \ \ \ \ \  \= $\text{{\sc Status}} \leftarrow \text{true}$. \\
            08. \> \> \> \> $H \leftarrow \text{{\sc Insert}}(H, e, G)$. \\
            09. \> \> \> \> {\sc While} there exists an edge $e \in H$ that is overfull w.r.t.~$H$: \\
            10. \> \> \> \> \ \ \ \ \ \ \ \ \ \= $H \leftarrow \text{{\sc Delete}}(H, e, G)$. \\
            11. \> \> {\sc If} $\text{{\sc Status}} = \text{false}$, {\sc Then} \\
            12. \> \> \> {\sc Break}.  \qquad \qquad \qquad \qquad // (This is the last round)  \\
            13. \> Let $U$ be the collection of edges in $E \setminus H$ that are underfull w.r.t.~$H$. Next,  consider any  set \\
             \>  $V_{small} \subseteq V$ s.t.~$\left\{ v \in V : \deg_v(U) \leq \frac{(1-\epsilon) \cdot (\Delta^*)^{\gamma}}{\epsilon} \right\} \subseteq V_{small} \subseteq \left\{ v \in V : \deg_v(U) \leq \frac{(1+\epsilon) \cdot (\Delta^*)^{\gamma}}{\epsilon} \right\}$, \\   
             \> and define $E_{small} := \{ (u, v) \in H \cup U : u, v \in V_{small}\}$. \\
            14. \> {\sc Return} $\mu(E_{small})$.
             \end{tabbing}
    \end{minipage}}}
    \caption{\label{alg:sample} SCHEMATIC-ALGO$(G = (V, E), \mu^*, m^*, \Delta^*, \gamma)$.}
\end{figure*}

\begin{lem}
\label{lem:bound:phases}
The algorithm in Figure~\ref{alg:sample} runs for at most $O(\beta^2 \cdot \mu(G))$ rounds.
\end{lem}

\begin{proof}
Consider the process where we keep changing the subset $H \subseteq E$ by repeatedly adding underfull edges and removing overfull edges. By Theorem~\ref{thm:local-moves}, this can continue for at most $O(\beta^2 \cdot \mu(G))$ iterations. The lemma follows since except the very last round, every other round in Figure~\ref{alg:sample} increases the number of iterations of this process by at least one (see step~(11) in Figure~\ref{alg:sample}). 
\end{proof}

\begin{lem}
\label{lem:deg:bound} Suppose that we are at the start of  round $k$ of the algorithm, for some $k \geq 1$. At this point in time, let  $H_k$ denote the state of the set $H$, and let $U_k$ denote the collection of edges in $E \setminus H_k$ that are underfull w.r.t.~$H_k$. If $|U_k| \geq \mu^{*} \cdot \left( \Delta^* \right)^{\gamma}$, then whp the algorithm does {\em not} terminate at the end of round $k$ (instead, it proceeds towards implementing round $k+1$).
\end{lem}

\begin{proof}
Throughout the proof, we condition on all the random choices made by the algorithm until the beginning of round $k$, and assume that $|U_k| \geq \mu^*  \left(\Delta^*\right)^{\gamma}$. During round $k$, the {\sc For} loop in step~(04) of Figure~\ref{alg:sample} runs for $(100 m^* \log n)/(\mu^* (\Delta^*)^{\gamma}) = L$ (say) iterations. Let $e_i \in E$ denote the edge that gets sampled during the $i^{th}$ iteration of this {\sc For} loop. For each $i \in [L]$, we have $\Pr[e_i \in U_k] = |U_k|/|E| \geq   \mu^*  (\Delta^*)^{\gamma}/m \geq \mu^* (\Delta^*)^{\gamma}/m^*$, since $m^* \geq m$ according to~(\ref{eq:parameters}). Let $X_i \in \{0, 1\}$ be an indicator random variable that is set to $1$ iff $e_i \in U_k$, and let $X = \sum_{i = 1}^L X_i$ denote the total number of times an edge from $U_k$ gets sampled during  round $k$. Note that the random variables $\{ X_i \}$ are mutually independent, and we have: $\mathbf{E}[X] = \sum_{i = 1}^L \mathbf{E}[X_i] = \sum_{i = 1}^L \Pr[e_i \in E_k] \geq |L| \cdot \mu^* (\Delta^*)^{\gamma}/m^* = 100 \log n$. From Chernoff bound, it follows that $X \geq 1$ whp.  In other words, whp at least one edge sampled during  round $k$ belongs to $U_k$. Consider the smallest index $i \in [L]$ such that $e_i \in U_k$; such an index exists whp. The algorithm  calls the subroutine {\sc Insert}$(H, e_i, G)$ during the $i^{th}$ iteration of the {\sc For} loop in round $k$, and sets $\text{{\sc Status}} \leftarrow \text{true}$  (see steps~(06) -- (08) in Figure~\ref{alg:sample}). This implies that the algorithm does {\em not} terminate at the end of round $k$ (see steps~(11)--(12) in Figure~\ref{alg:sample}).
\end{proof}

\begin{cor}
\label{cor:deg:bound}
At the end of the last round in Figure~\ref{alg:sample}, we have $|U| < \mu^* \cdot (\Delta^*)^{\gamma}$ whp.
\end{cor}

\begin{proof}
Follows from Lemma~\ref{lem:bound:phases} and Lemma~\ref{lem:deg:bound}.
\end{proof}

\begin{lem}
\label{lem:approx}
At step~(14) in Figure~\ref{alg:sample}, we have $\mu(G) \leq (3/2+\epsilon) \cdot \mu(E_{small}) + \Theta(\epsilon) \cdot \mu^*$, whp.
\end{lem}

\begin{proof}
Initially, the algorithm starts by setting $H = \emptyset$, and hence $H$ is an EDBS of $G = (V, E)$ at this point. Steps~(09)--(10) in Figure~\ref{alg:sample} ensures that the set $H$ continues to remain an EDBS of $G$ throughout the duration of the algorithm. Since $U \subseteq E \setminus H$ is the collection of underfull edges w.r.t.~$H$ at the end of the {\sc Repeat} loop in Figure~\ref{alg:sample},  Theorem~\ref{thm:EDCS} implies that: 
\begin{equation} 
\label{eq:approx}
\mu(G) \leq (3/2+\epsilon) \cdot \mu(U \cup H)
\end{equation}
Henceforth, condition on the high probability event that $|U| < \mu^*  (\Delta^*)^{\gamma}$ (see Corollary~\ref{cor:deg:bound}). Since each node $v \in  V \setminus V_{small}$ is incident upon at least $\frac{(1-\epsilon)(\Delta^*)^{\gamma}}{\epsilon}$ edges from $U$, we infer that: 
$$|V \setminus V_{small}| \leq \frac{2 \cdot  |U|}{\frac{(1-\epsilon)(\Delta^*)^{\gamma}}{\epsilon}} < \frac{2 \epsilon \mu^*}{(1-\epsilon)} < 3\epsilon \mu^*.$$ Accordingly, if we remove all the edges in $H \cup U$ that share an endpoint with $V \setminus V_{small}$, then the size of the maximum matching in $H \cup U$ can decrease by at most an additive $3 \epsilon \mu^*$. To see why this is the case, fix any maximum matching $M$ in $H \cup U$, and note that each time we remove a node (along with its incident edges) from $H \cup U$, the number of edges in $M$ decreases by at most one. Finally, note that once we have removed all the edges incident on $V \setminus V_{small}$ from $H \cup U$, we end up with the subgraph $E_{small}$. Hence, we have:
\begin{equation}
\label{eq:approx:2}
\mu(H \cup U) \leq \mu(E_{small}) + 3 \epsilon \mu^*.
\end{equation}
The lemma now follows from~(\ref{eq:approx}) and~(\ref{eq:approx:2}).
\end{proof}

\begin{lem}
\label{lem:maxdeg:schematic} The graph $G_{small} := (V_{small}, E_{small})$ has maximum degree $\Delta_{G_{small}} = O_{\epsilon}\left( (\Delta^*)^{\gamma}\right)$.
\end{lem}

\begin{proof}
Consider any node $v \in V_{small}$. By definition, we have $\deg_v(U) = O_{\epsilon}\left((\Delta^*)^{\gamma}\right)$. On the other hand, since $H$ is an EDBS of $G$, we have $\deg_e(H) \leq \beta$ for all edges $e \in H$, and hence $\deg_u(H) \leq \beta$ for all nodes $u \in V$. In particular, this means that $\deg_v(H) \leq \beta$, and hence $\deg_v(E_{small}) \leq \deg_v(H) + \deg_v(U) = O_{\epsilon}\left(\beta + (\Delta^*)^{\gamma}\right) = O_{\epsilon}\left((\Delta^*)^{\gamma}\right)$. This implies the lemma.
\end{proof}

\section{Implementation}

\label{sec:implementation}

We now show how to implement the schematic algorithm from Section~\ref{sec:schematic} in sublinear settings, under both adjacency-matrix and adjacency-list queries. To highlight the main ideas and for simplicity of exposition, here we only focus on bounding the query-complexities of our sublinear algorithms. In Appendix~\ref{sec:appendix:runtimeyoshida}, we explain how our algorithms can be efficiently implemented, in such a way that their running times have similar asymptotic bounds as their query complexities.

\subsection{$(3/2, \epsilon n)$-Approximation with Adjacency-Matrix Queries}
\label{sec:adj-matrix}

The main results in this section are summarized  in the theorem and the corollary below.

\begin{thm}
\label{th:main:adj-matrix}
There is an algorithm which, given an input graph $G = (V, E)$,   returns a $(3/2, \epsilon n)$-approximation to the value of $\mu(G)$  after  $\tilde{O}_{\epsilon}\left(n^{2-\Theta(\epsilon^2)}\right)$  adjacency-matrix queries in $G$,   whp.
\end{thm}

\begin{cor}
\label{cor:main:adj-matrix}
Suppose that there is an algorithm $\mathcal{A}^*$ which, given an input graph $G$ with maximum degree $\Delta$, whp returns a $(1, \epsilon n)$-approximation to the value of $\mu(G)$ after $\tilde{O}_{\epsilon}(n \Delta^{q})$ adjacency-matrix queries in $G$, where $q \geq 0$. Then there exists another adjacency-matrix query based algorithm which has the same approximation guarantee as in Theorem~\ref{th:main:adj-matrix}, and query complexity   $\tilde{O}_{\epsilon}\left(n^{2-1/(1+q)}\right)$.
\end{cor}

We devote the rest of this section to proving Theorem~\ref{th:main:adj-matrix} and Corollary~\ref{cor:main:adj-matrix}. We   implement the schematic algorithm described in Figure~\ref{alg:sample}, in a setting where we can access the edges of $G$ only via adjacency-matrix queries. Before proceeding any further, we show how to handle two basic primitives in the adjacency-matrix query model: (a) how to estimate the value of $m = |E|$, and (b) how to sample an edge uniformly at random from $E$. The tools for implementing these primitives are respectively captured in Lemma~\ref{lem:primitive:1} and Lemma~\ref{lem:primitive:2}; their proofs are deferred to Appendix~\ref{appendix:missingproofs:approx-matching}.
 
 \begin{lem}
 \label{lem:primitive:1} Suppose that we sample, independently and uniformly at random (with repetitions), $S = (100/\epsilon^3) \cdot n \log n$ many pairs of nodes from $V \times V$. For each $i \in [S]$, let $X_i \in \{0, 1\}$ be an indicator random variable that is set to $1$ iff the $i^{th}$ pair of nodes sampled in this manner are connected via an edge in $G$. Let $X := \sum_{i=1}^S X_i$. Define
 $$\hat{m} := (1+\epsilon) \cdot \frac{X n^2}{S} +\epsilon n.$$ Then we have  $m \leq \hat{m} \leq  (1+\epsilon) \cdot m + 2 \cdot \epsilon n$, whp.
 \end{lem}

\begin{lem}
\label{lem:primitive:2} Suppose that we keep sampling, independently and uniformly at random (with repetitions), unordered pairs of nodes from  $V \times V$, until we find an edge $e \in E$. Then:
\begin{itemize}
    \item (i) The edge $e$ we find is a uniformly random sample from the set $E$.
    \item (ii) Whp, this random process stops after sampling $\tilde{O}(n^2/m)$ pairs of nodes from  $V \times V$.
\end{itemize}
\end{lem}

\medskip
\noindent 
Our algorithm under adjacency-matrix queries consists of the following four steps.

\medskip
\noindent {\bf Step 1: Preprocessing.} We first invoke Lemma~\ref{lem:primitive:1} to get an estimate $\hat{m}$ of the value of $m$, using $S = \tilde{O}_{\epsilon}(n)$ adjacency-matrix queries. From this point onward, we condition on the high probability event guaranteed by Lemma~\ref{lem:primitive:1}. If $\hat{m} \leq 4 \epsilon n$, then we infer that $m = O(\epsilon n)$, and so our algorithm stops execution and returns the value $0$ as an estimate of $\mu(G)$, which is trivially a $(1, \epsilon n)$-approximation to $\mu(G)$ since $\mu(G) \leq m = O(\epsilon n)$. Accordingly, for the rest of this section, we assume that $\hat{m} > 4 \epsilon n$. In this case, by Lemma~\ref{lem:primitive:1} we have:
\begin{equation}
    \label{eq:estimate}
    m \leq \hat{m} \leq \Theta(m).
\end{equation}
At this point, we set $\mu^* \leftarrow n$, $m^* \leftarrow \hat{m}$, $\Delta^* \leftarrow n$ and $\gamma \leftarrow c \cdot \epsilon^2$ for a sufficiently small constant $c > 0$. We next show how to implement a call to SCHEMATIC-ALGO$(G, \mu^*, m^*, \Delta^*, \gamma)$ as  in Figure~\ref{alg:sample}.

\medskip
\noindent {\bf Step 2: Obtaining the EDBS $H$.} The goal here is to  implement steps (01)--(12) of Figure~\ref{alg:sample}. Note that in a given ``round'' (an iteration of the {\sc Repeat} loop in Figure~\ref{alg:sample}), we need to sample $\Theta(m \log n/n  ^{1+\gamma})$ edges from $E$ uniformly and independently at random. To obtain each of these samples, we invoke Lemma~\ref{lem:primitive:2}. This requires us to make $\tilde{O}(n^2/m)$ adjacency-matrix queries per sample, whp. Once we have the sampled edges at our disposal, we can perform all the remaining computation in that round without any further overhead in our query complexity. Thus, to implement one round we  make a total of $\tilde{O}(m/n^{1+\gamma}) \cdot \tilde{O}(n^2/m) = \tilde{O}(n^{1-\gamma})$ adjacency-matrix queries in $G$, whp. By Lemma~\ref{lem:bound:phases}, there are at most $O(\beta^2 \cdot \mu(G)) = O_{\epsilon}(n)$ rounds. Hence, to implement all these rounds, whp overall we  make  $\tilde{O}_{\epsilon}\left(n \cdot n^{1-\gamma} \right) = \tilde{O}_{\eps}(n^{2-\gamma})$ adjacency-matrix queries in $G$.

\medskip
Note that the end of Step 2, our algorithm explicitly stores the EDBS $H$. The main challenge for our adjacency-matrix query based algorithm now is to compute an approximation to the value of $\mu(H \cup U)$, given that it does not explicitly store the set of underfull edges $U$.

\medskip
\noindent {\bf Step 3: Identifying the set $V_{small}$.} For each node $v \in V$, we now decide whether or not to include it in the set $V_{small}$ in the following manner.
\begin{itemize}
    \item Sample $K = (100/\epsilon) \cdot  n^{1-\gamma} \log n$ nodes $v_1, \ldots, v_K$ from the set $V$, uniformly and independently at random (with repetitions). For each $i \in [K]$, check whether $(v, v_i) \in U$: by first making an adjacency-matrix query in $G$ to confirm whether $(v, v_i) \in E$, and if yes, then by further verifying whether $(v, v_i) \in E \setminus H$ and $\deg_v(H) + \deg_{v_i}(H) \leq (1-\epsilon) \beta$ (these last two conditions can be checked without making any further adjacency-matrix queries in $G$, since we explicitly store all the edges in $H$). Let $X_{i, v} \in \{0, 1\}$ be an indicator random variable that is set to $1$ iff $(v, v_i) \in E$. Let  $X_v := \sum_{i=1}^K X_{i, v}$. We include the node $v$ into $V_{small}$ iff $X_v \leq (100/\epsilon^2) \cdot \log n$.
\end{itemize}
 The total number of adjacency-matrix queries in $G$ made during the above procedure is  $n \cdot K = \tilde{O}_{\epsilon}(n^{2-\gamma})$. We now show that this procedure is consistent with  step~(13) of Figure~\ref{alg:sample}.

\begin{claim}
\label{cl:degree:adj-matrix}
The set $V_{small}$, as constructed above, whp satisfies the condition: 
$$\left\{ v \in V : \deg_v(U) \leq \frac{(1-\epsilon) \cdot n^{\gamma}}{\epsilon} \right\} \subseteq V_{small} \subseteq \left\{ v \in V : \deg_v(U) \leq \frac{(1+\epsilon) \cdot n^{\gamma}}{\epsilon} \right\}.$$
\end{claim}

\begin{proof}
Fix any node $v \in V$. The random variables $\{X_{i, v}\}$ are mutually independent, and we have $\mathbf{E}[X_v] = \sum_{i=1}^K \mathbf{E}[X_{i, v}] = \sum_{i=1}^K \Pr[X_{i, v} = 1] = K \cdot \deg_v(U)/n = (100/\epsilon) \cdot n^{1-\gamma} \log n \cdot \deg_v(U)/n = (100 /\epsilon) \cdot n^{-\gamma} \log n \cdot \deg_v(U)$.  Thus, applying Chernoff bounds, we infer that:
\begin{obs}
\label{ob:first:adj-matrix}
If  $\deg_v(U) \leq \frac{(1-\epsilon) \cdot n^{\gamma}}{\epsilon}$, then  whp $v \in V_{small}$.  On the other hand, if  $\deg_v(U) > \frac{(1+\epsilon) \cdot n^{\gamma}}{\epsilon}$, then  whp $v \notin V_{small}$.
\end{obs}
The claim now follows from Observation~\ref{ob:first:adj-matrix} and a union bound over all the nodes $v \in V$.
\end{proof}

\medskip
To summarize, at the end of Step 3, we explicitly store the EDBS $H$ and  the node-set $V_{small}$. We, however, only have adjacency-matrix query access to the edges in $G_{small} := (V_{small}, E_{small})$.  Indeed, suppose that we are given a pair of nodes $u, v \in V_{small}$, and we are asked to determine whether $(u, v) \in E_{small}$. To perform this task, we first make an adjacency-matrix query in $G$ about the existence of the edge $(u, v)$ in $E$. If the response  tells us that $(u, v) \in E$, then we further check whether  the following condition holds: either $(u, v) \in H$, or  $\deg_u(H) + \deg_v(H) < (1-\epsilon)\beta$. If the answer is yes, then we conclude that $(u, v) \in E_{small}$. Otherwise, we conclude that $(u, v) \notin E_{small}$.

\medskip
\noindent {\bf Step 4: Computing a $(1+\epsilon)$-approximate estimate of $\mu(E_{small})$.}  We now run the algorithm from Theorem~\ref{thm:Yoshida} on  $G_{small} := (V_{small}, E_{small})$, and output the estimate $\alpha$ returned by this algorithm. The key challenge here is that Theorem~\ref{thm:Yoshida} requires adjacency-list query access to $G_{small}$, but we only have adjacency-matrix query access to $G$. We overcome this challenge  as follows. 

Suppose that  the algorithm from Theorem~\ref{thm:Yoshida} makes an adjacency-list query $Q$, asking for the $i^{th}$ edge incident on a given node $v \in V_{small}$ in $G_{small}$. To answer this  query $Q$, we make $(n-1)$ adjacency-matrix queries in $G$ -- to collect all the edges incident on $v$ in $G$ -- and then identify the subset of these edges that belong to $G_{small}$ (based on the contents of  the sets $H$ and $V_{small}$, which we store explicitly). And once we have identified this subset, we can trivially answer the original  query  $Q$.  To summarize, we can implement each adjacency-list query in $G_{small}$ by making $(n-1)$ adjacency-matrix queries in $G$.

By Lemma~\ref{lem:maxdeg:schematic} and Claim~\ref{cl:degree:adj-matrix}, whp the  graph $G_{small}$ has maximum-degree $\Delta_{G_{small}} = O_{\epsilon}(n^{\gamma})$. Accordingly,  it follows that in order to run the algorithm from Theorem~\ref{thm:Yoshida} on $G_{small}$, whp we make at most $\tilde{O}_{\epsilon}\left(n \cdot (\Delta_{G_{small}})^{\Theta(1/\epsilon^2)}\right) = \tilde{O}_{\epsilon}\left(n^{1+\Theta(\gamma/\epsilon^2)}\right)$ adjacency-matrix queries in $G$.

\medskip
\noindent {\bf Approximation guarantee:} 
As discussed above, our algorithm returns an estimate $\alpha$ such that $\alpha \leq \mu(E_{small}) \leq (1+\epsilon) \cdot \alpha$. Since we have $\mu^* = n$ (see Step 1 above), Lemma~\ref{lem:approx} implies that:
\begin{equation}
\label{eq:approx:main:adj-matrix}
\alpha \leq \mu(E_{small}) \leq \mu(G) \leq  (3/2 + \Theta(\epsilon)) \cdot \alpha + \Theta(\epsilon n) = (3/2) \cdot \alpha + \Theta(\epsilon n).
\end{equation}

\medskip
\noindent {\bf Proof of Theorem~\ref{th:main:adj-matrix}.}
The approximation guarantee follows from~(\ref{eq:approx:main:adj-matrix}). Next, recall that Steps 1, 2, 3, 4 respectively makes at most  $\tilde{O}_{\eps}(n)$, $\tilde{O}_{\eps}(n^{2-\gamma})$, $\tilde{O}_{\eps}(n^{2-\gamma})$ and $\tilde{O}_{\eps}\left( n^{1+\Theta(\gamma/\epsilon^2)}\right)$  adjacency-matrix queries in $G$, whp. Hence, the total  query complexity of our algorithm is at most:
\begin{eqnarray*}
 \tilde{O}_{\eps}(n) + \tilde{O}_{\eps}(n^{2-\gamma}) +\tilde{O}_{\eps}(n^{2-\gamma}) + \tilde{O}_{\eps}\left( n^{1+\Theta(\gamma/\epsilon^2)}\right) =  \tilde{O}_{\epsilon}\left(  n^{2-\Theta(\epsilon^2)} \right).
\end{eqnarray*}
The last equality holds since  $\gamma := c \cdot \epsilon^2$ for a sufficiently small constant $c > 0$ (see Step 1 above). \qed 

\medskip
\noindent {\bf Proof of Corollary~\ref{cor:main:adj-matrix}.} Steps 1, 2 and 3 of the algorithm remains the same as before. In Step 4, however, we run the algorithm $\mathcal{A}^*$ (see the statement of the corollary) on  $G_{small}$. The query-complexity of implementing Step 4 now becomes $\tilde{O}_{\epsilon}\left( n \cdot \left( \Delta_{G_{small}} \right)^q \right) = \tilde{O}_{\epsilon} \left( n^{1+q\gamma}\right)$. Hence, if we set $\gamma := 1/(1+q)$, then the total query complexity of the algorithm, over all the four steps, becomes:
\begin{eqnarray*}
 \tilde{O}_{\eps}(n) + \tilde{O}_{\eps}(n^{2-\gamma}) +\tilde{O}_{\eps}(n^{2-\gamma}) + \tilde{O}_{\eps}\left( n^{1+q\gamma}\right) =  \tilde{O}_{\epsilon}\left(  n^{2-1/(1+q)} \right). \qed
\end{eqnarray*}

\subsection{$(3/2 + \epsilon)$-Approximation with Adjacency-List Queries}
\label{sec:adj-list}

The main result in this section is summarized in the theorem below.

\begin{thm}
\label{th:main:adj-list}
There is an algorithm which, given an input graph $G = (V, E)$, returns a $(3/2+\epsilon)$-approximation to the value of $\mu(G)$ after
 $\tilde{O}_{\epsilon}\left(n \cdot \Delta^{1-\Theta(\epsilon^2)}\right)$  adjacency-list queries in $G$,  whp.\footnote{Here, $\Delta$ denotes the maximum degree of any node in $G$.}
\end{thm}

We devote the rest of this section to proving Theorem~\ref{th:main:adj-list}. We  implement the schematic algorithm described in Figure~\ref{alg:sample}, in a setting where we have access to the input graph only via adjacency-list queries. Specifically, the algorithm consists of the following five steps.

\medskip
\noindent {\bf Step 1: Preprocessing.} For each  $v \in V$, we compute its degree $\deg_v(E)$ in  $G$ by doing a binary search on the interval $[0, n]$, using $O(\log n)$ adjacency-list queries. Thus, at the end of $\tilde{O}(n)$ adjacency-list queries, we know the degree of each node in $G$, along with the values of $m, d$ and $\Delta$. 

\medskip
This means that from now on we can  sample a uniformly random edge from $G$ using a single adjacency-list query: For this, all we need to do is sample a node $v$ from the distribution $\mathcal{D}$, which places a probability mass of $\deg_x(E)/(2 \cdot |E|)$ on each node $x \in V$, and then sample an index $i \in [\deg_v(E)]$ uniformly at random, and make an adjacency-list query to return the $i^{th}$ edge incident on $v$. This observation will be used in Step 3 of our algorithm below.

\medskip
\noindent {\bf Step 2: Obtaining a coarse approximation of $\mu(G)$.} We call the algorithm from Theorem~\ref{thm:Soheil}, which returns an estimate $\lambda$ such that $\lambda \leq \mu(G) \leq (2+\epsilon) \cdot \lambda$, using $\tilde{O}_{\eps}(n)$ adjacency-list queries. We set $\mu^* \leftarrow \lambda$, $m^* \leftarrow m$, $\Delta^* \leftarrow \Delta$ and $\gamma \leftarrow c \cdot \epsilon^2$ for a sufficiently small constant $c > 0$, and we next show how to implement a call to SCHEMATIC-ALGO$(G, \mu^*, m^*, \Delta^*, \gamma)$, as described in Figure~\ref{alg:sample}.

\medskip
\noindent {\bf Step 3: Obtaining the EDBS $H$.} The goal here is to implement steps (01)--(12) of Figure~\ref{alg:sample}. Note that in a given ``round'' (an iteration of the {\sc Repeat} loop in Figure~\ref{alg:sample}), we need to sample $\Theta(m \log n/(\lambda  \Delta^{\gamma}))$ edges from $E$ uniformly and independently at random, and once we have the sampled edges at our disposal, we can perform all the remaining computation in that round without any further overhead in our query complexity. We have, however, already discussed in Step 1 that by making a single adjacency-list query, we can sample a uniformly random edge from $G$. Thus,  a given round in Figure~\ref{alg:sample} requires us to make $\Theta(m \log n/(\lambda \Delta^{\gamma}))$ adjacency-list queries. By Lemma~\ref{lem:bound:phases}, there are at most $O(\beta^2 \cdot \mu(G)) = O(\beta^2 \lambda)$ rounds. Hence, to implement all these rounds, we  make at most $\Theta\left(\beta^2 \lambda \cdot (m \log n)/(\lambda \Delta^{\gamma}) \right) = \tilde{O}_{\eps}(m/\Delta^{\gamma}) = \tilde{O}_{\epsilon}(n \cdot \Delta^{1-\gamma})$ adjacency-list queries.

\medskip
At the end of Step 3, we only explicitly store the EDBS $H$. The main challenge for our adjacency-list query based algorithm now is to compute an approximation to the value of $\mu(H \cup U)$, given that it does not explicitly store the set of underfull edges $U$.

\medskip
\noindent {\bf Step 4: Identifying the set $V_{small}$.} For each node $v \in V$, we sample $K = (100/\epsilon) \cdot \Delta^{1-\gamma} \log n$ edges in $G$ incident on $v$, uniformly and independently at random (with repetitions). We need to make one adjacency-list query for each of these samples, which leads to a total of $\tilde{O}_{\epsilon}( \Delta^{1-\gamma})$ adjacency-list queries per node, and hence $\tilde{O}_{\eps}(n \cdot \Delta^{1-\gamma})$ adjacency-list queries to deal with all the  $n$ nodes in $G$. For each $i \in [K]$, let $(v, u_i) \in E$ denote the $i^{th}$ sample for $v \in V$, and let $X_{i, v} \in \{0, 1\}$ be an indicator random variable that is set to $1$ iff $(v, u_i) \in U$. Note that once we have retrieved the sampled edge $(v, u_i) \in E$, it is easy for us to check whether it belongs to the set $U$ (by verifying whether $(v, u_i) \in E \setminus H$ and whether $\deg_v(H) + \deg_{u_i}(H) \leq (1-\epsilon) \beta$) without making any further adjacency-list queries in $G$, since we explicitly store all the edges in $H$. Let  $X_v := \sum_{i=1}^K X_{i, v}$. We construct the set $V_{small}$ as follows: $V_{small} := \{ v \in V : X_v \leq (100/\epsilon^2) \cdot \log n\}$. We now show that this is consistent with the way $V_{small}$ is defined in step~(13) of Figure~\ref{alg:sample}.

\begin{claim}
\label{cl:degree}
The set $V_{small}$, as constructed above, whp satisfies the condition: 
$$\left\{ v \in V : \deg_v(U) \leq \frac{(1-\epsilon) \cdot \Delta^{\gamma}}{\epsilon} \right\} \subseteq V_{small} \subseteq \left\{ v \in V : \deg_v(U) \leq \frac{(1+\epsilon) \cdot \Delta^{\gamma}}{\epsilon} \right\}.$$
\end{claim}

\begin{proof}
Fix any node $v \in V$. The random variables $\{X_{i, v}\}$ are mutually independent, and we have $\mathbf{E}[X_v] = \sum_{i=1}^K \mathbf{E}[X_{i, v}] = \sum_{i=1}^K \Pr[X_{i, v} = 1] = K \cdot \deg_v(U)/\deg_v(E) = (100/\epsilon) \cdot \Delta^{1-\gamma} \log n \cdot \deg_v(U)/\deg_v(E) \geq (100/\epsilon) \cdot \Delta^{-\gamma} \log n \cdot \deg_v(U)$. Thus, applying Chernoff bounds, we get:
\begin{obs}
\label{ob:first}
If  $\deg_v(U) \leq \frac{(1-\epsilon) \cdot \Delta^{\gamma}}{\epsilon}$, then  whp $v \in V_{small}$.  On the other hand, if  $\deg_v(U) > \frac{(1+\epsilon) \cdot \Delta^{\gamma}}{\epsilon}$, then  whp $v \notin V_{small}$.
\end{obs}
The claim now follows from Observation~\ref{ob:first} and a union bound over all the nodes $v \in V$.
\end{proof}

\medskip
To summarize, at the end of Step 4, we explicitly store the EDBS $H$ and  the node-set $V_{small}$, but {\em not} the edge-set $E_{small}$ (see step~(13) in Figure~\ref{alg:sample}). However, given a pair of nodes $u, v \in V_{small}$, along with a guarantee (which we are in no position to verify) that $u$ and $v$ are connected via an edge in $E$, we can decide whether or not $(u, v) \in E_{small}$ based on the contents of the set $H$.

\medskip
\noindent {\bf Step 5: Computing a $(1+\epsilon)$-approximate estimate of $\mu(E_{small})$.} We now run the algorithm from Theorem~\ref{thm:Yoshida} on $G_{small} := (V_{small}, E_{small})$, and output the estimate $\alpha$ returned by this algorithm. The key challenge here is that Theorem~\ref{thm:Yoshida} requires adjacency-list query access to $G_{small}$, but we only have adjacency-list query access to $G$. We overcome this challenge as follows.

  Suppose that  this algorithm makes an adjacency-list query $Q$, asking for the $i^{th}$ edge incident on a given node $v \in V_{small}$ in $G_{small}$. To answer this  query $Q$, we make $\deg_v(E)$ many adjacency-list queries in $G$ -- to collect all the edges incident on $v$ in $G$ -- and then identify the subset of these edges that belong to $G_{small}$ (based on the contents of  $H$ and $V_{small}$, which we explicitly store). Once we have identified this subset, we can trivially answer the original  query  $Q$.  To summarize, we can implement each adjacency-list query in $G_{small}$ by making at most $\Delta$ adjacency-list queries in $G$.

By Lemma~\ref{lem:maxdeg:schematic} and Claim~\ref{cl:degree} , whp the  graph $G_{small}$ has maximum-degree $\Delta_{G_{small}} = O_{\epsilon}(\Delta^{\gamma})$. Accordingly, it follows that in order to run the algorithm from Theorem~\ref{thm:Yoshida} on $G_{small}$, we make $\tilde{O}_{\epsilon}\left(\Delta \cdot (\Delta_{G_{small}})^{\Theta(1/\epsilon^2)}\right) = \tilde{O}_{\epsilon}\left( \Delta^{(1+\Theta(\gamma/\epsilon^2))} \right)$ adjacency-list queries in $G$, whp.

\medskip
\noindent {\bf Approximation Guarantee:}
As discussed above, our algorithm returns an estimate $\alpha$ s.t.~$\alpha \leq \mu(E_{small}) \leq (1+\epsilon) \cdot \alpha$. Since $\lambda = \mu^* \leq \mu(G) \leq (2+\epsilon) \cdot \mu^*$ (see Step 2), Lemma~\ref{lem:approx} implies that:
\begin{equation}
\label{eq:approx:main}
\alpha \leq \mu(E_{small}) \leq \mu(G) \leq  (3/2 + \Theta(\epsilon)) \cdot \alpha.
\end{equation}

\medskip
\noindent {\bf Proof of Theorem~\ref{th:main:adj-list}.}
The approximation guarantee follows from~(\ref{eq:approx:main}). Next, recall that Steps 1, 2, 3, 4, 5 respectively makes  $\tilde{O}(n)$, $\tilde{O}_{\eps}(n)$, $\tilde{O}_{\eps}(n \cdot \Delta^{1-\gamma})$, $\tilde{O}_{\eps}(n \cdot \Delta^{1-\gamma})$ and $\tilde{O}_{\eps}\left(\Delta^{1+\Theta(\gamma/\epsilon^2)}\right)$  adjacency-list queries in $G$, whp. Hence, the total query complexity of the algorithm is at most:
\begin{eqnarray*}
\tilde{O}(n) + \tilde{O}_{\eps}(n) + \tilde{O}_{\eps}(n \cdot \Delta^{1-\gamma}) +\tilde{O}_{\eps}(n \cdot \Delta^{1-\gamma}) + \tilde{O}_{\eps}\left( \Delta^{1+\Theta(\gamma/\epsilon^2)}\right)    = \tilde{O}_{\epsilon}\left( n \cdot \Delta^{1-\Theta(\epsilon^2)} \right).
\end{eqnarray*}
This holds because $\Delta \leq n$, and $\gamma := c \cdot \epsilon^2$ for a sufficiently small constant $c > 0$ (see Step 2). \qed

\subsection{$(3/2, \epsilon n)$-Approximation with Adjacency-List Queries}
\label{sec:adj-list:hybrid}

In Section~\ref{sec:adj-list}, we showed how to obtain a $(3/2+\epsilon)$-approximation to $\mu(G)$ with $\tilde{O}_{\epsilon} (n \cdot \Delta^{1-\Theta(\epsilon^2)})$ adjacency-list queries. Note that this query complexity is sublinear in $m = |E|$ only if most of the nodes in $G$ have degree close to $\Delta$. In this section, we address this issue by designing an algorithm whose query complexity is always sublinear in $m$, at the cost of incurring an additive $\epsilon n$ slack in the approximation guarantee. The main result in this section is summarized in the theorem below.

\begin{thm}
\label{th:main:adj-list:hybrid}
There exists an algorithm which, on input graph $G = (V, E)$,  returns a $(3/2, \epsilon n)$-approximation to the value of $\mu(G)$ after
$\tilde{O}_{\epsilon}\left(n \cdot \left( 1+ d^{1-\Theta(\epsilon^2)}\right)\right)$  adjacency-list queries to $G$, whp.\footnote{Here, $d$ denotes the average degree in $G$.}
\end{thm}

Steps 1, 2, 3 of our algorithm for Theorem~\ref{th:main:adj-list:hybrid} remains almost the same as Steps 1, 2, 3 in Section~\ref{sec:adj-list}; the only difference being that in Step 2 we set $\Delta^* \leftarrow d$ (instead of setting $\Delta^* \leftarrow \Delta$ as in Section~\ref{sec:adj-list}). After the end of Step 3, we diverge from the algorithm in Section~\ref{sec:adj-list}, for the following reason. For each node $v \in V$, Step 4 in Section~\ref{sec:adj-list} decides whether or not it belongs to $V_{small}$ by sampling $\tilde{\Theta}_{\epsilon}( \Delta^{1-\gamma})$ edges incident on $v$ in $G$, and then checking how many of these samples belong to $H$. This leads to an overall query complexity of $\tilde{O}_{\epsilon}(n \cdot \Delta^{1-\gamma})$ for Step 4. If we instead wish to get a similar query complexity in terms of $d$, then we can afford to sample at most $\tilde{\Theta}_{\epsilon}( d^{1-\gamma})$ incident edges per node. However, if a node $v \in V$ has  $\deg_v(E) \gg d$, then $\tilde{\Theta}_{\epsilon}( d^{1-\gamma})$ samples will never suffice to determine whether $\deg_v(H) = O(d^{\gamma}/\epsilon)$. Note that this difficulty does not arise in Section~\ref{sec:adj-list}, because every node in $G$ has degree at most $\Delta$ (in contrast to the current section, where some nodes can have degree $\gg d$). To overcome this difficulty, we prune the high degree nodes out of $G$, and work with the remaining subgraph while implementing Step 4 and Step 5, following the template of Section~\ref{sec:adj-list}. We defer the complete algorithm and its analysis to Appendix~\ref{appendix:missingproofs:approx-matching}.

\section{Returning a Matching vs Approximating the Value of $\mu(G)$}

\label{sec:dichotomy}

It is well-known that no algorithm can return the edges of a $O(1)$-approximate maximum matching in an input graph $G$ using only sublinear number of queries. In addition, it remains a big open question to design a sublinear algorithm that returns a $(1, \epsilon n)$-approximation (or a $(1+\epsilon)$-approximation) to the value of $\mu(G)$. As summarized in the theorem below, an interesting corollary of our  analysis from the previous sections is that on any given graph, we can achieve one of these two goals.   

\begin{thm}
\label{thm:dichotomy} There is an algorithm for each of the following tasks, on an input graph $G = (V, E)$.
\begin{itemize}
    \item (i) Either return a matching $M \subseteq E$ of size $|M| = \Omega(\epsilon \cdot \mu(G))$ or return a $(1, \epsilon n)$-approximation to the value of $\mu(G)$, by making $\tilde{O}_{\epsilon}\left( n^{2-\Theta(\epsilon^2)}\right)$ adjacency-matrix queries, whp.
    \item (ii) Either return a matching $M \subseteq E$ of size $|M| = \Omega(\epsilon \cdot \mu(G))$ or return a $(1+ \epsilon )$-approximation to the value of $\mu(G)$, by making $\tilde{O}_{\epsilon}\left( n \cdot \Delta^{1-\Theta(\epsilon^2)}\right)$ adjacency-list queries, whp.
    \item (iii) Either return a matching $M \subseteq E$ of size $|M| = \Omega(\epsilon \cdot \mu(G))$ or return a $(1, \epsilon n)$-approximation to the value of $\mu(G)$, by making $\tilde{O}_{\epsilon}\left(n \cdot \left(1+d^{1-\Theta(\epsilon^2)}\right)\right)$ adjacency-list queries, whp.
\end{itemize}
\end{thm}

\begin{proof}(Sketch) Recall that all the three sublinear algorithms described in Section~\ref{sec:implementation} are based on the template established in Figure~\ref{alg:sample}. Furthermore, all the three sublinear algorithms satisfy the following properties: (i) they explicitly store the contents of the EDBS $H$ of $G$ (and hence can output the set of edges belonging to a maximum matching in $H$, if required), and (ii) they return a near-optimal approximation to the value of $\mu(H \cup U)$. The theorem now follows from Lemma~\ref{lem:EDCS:extremecase}.
\begin{lem}

\label{lem:EDCS:extremecase}

Let $H \subseteq E$ be an EDBS of  $G = (V, E)$, and let $U \subseteq E \setminus H$ denote the collection of edges that are underfull w.r.t.~$H$. Then, either $\mu(G) \leq 
(1/\Theta(\epsilon)) \cdot \mu(H)$ or    $\mu(G) \leq  (1 + \Theta(\epsilon)) \cdot  \mu(H \cup U)$.
\end{lem}
The proof of Lemma~\ref{lem:EDCS:extremecase} appears in Section~\ref{sec:lem:EDCS:extremecase}. 
\end{proof}

\subsection{Proof of Lemma~\ref{lem:EDCS:extremecase}}
\label{sec:lem:EDCS:extremecase}

The proof follows the ideas of \cite{AssadiB19}. We will first prove the lemma for bipartite graphs.

\begin{claim}
\label{cl:dichotomy:100}
Consider a bipartite graph $G = (L \cup R,E)$, and an EDBS $H \subseteq E$ of $G$. Let $U \subseteq E \setminus H$ be the collection of edges that are underfull w.r.t.~$H$. Then, either $\mu(G) \leq (1/\epsilon) \cdot \mu(H)$ or $\mu(G) \leq (1 + \Theta(\epsilon)) \cdot \mu(H \cup U)$.

\end{claim}

\begin{proof}
Applying Hall's theorem on the subgraph $H$, we infer that there must exist a subset of nodes $A \subseteq R$ such that $|A| - |N_H(A)| = |R| - \mu(H)$, where $N_H(A) := \{ u \in L : (u, v) \in H \text{ for some } v \in A\}$ denotes the set of neighbors of $A$ in $H$. Let $B := N_H(A)$, $\overline{B} := L \setminus B$, and $\overline{A} := R \setminus A$. This implies that $|A| - |B| = |R| - \mu(H)$, and hence $|\overline{A}| + |B| = \mu(H)$.

Note that there must exist a matching  $M \subseteq E \setminus H$ of size $|M| \geq \mu(G) - \mu(H)$ between the vertices of $\overline{B}$ and $A$, for otherwise $A$ would be a Hall's witness set showing that the maximum matching size of $G$ is smaller then $\mu(G)$, leading to a contradiction. Let $S \subseteq A \cup \overline{B}$ denote the set of endpoints of the edges of $M$. Since $N_H(S) \subseteq  \overline{A} \cup B$, we have $|N_H(S)| \leq |\overline{A}| + |B| = \mu(H)$.

For sake of contradiction, suppose that  $\mu(H) < \epsilon \cdot \mu(G)$ and $\mu(H \cup U) < (1-5\epsilon)\cdot \mu(G)$. Then, we have $|M| \geq \mu(G) - \mu(H) \geq (1-\epsilon) \cdot \mu(G)$. Since $\mu(H \cup U) < (1- 5 \epsilon) \cdot \mu(G)$ and $|M| \geq (1-\epsilon) \cdot \mu(G)$, it follows that at least $4 \epsilon \cdot \mu(G)$ edges of $M$ are not in $H \cup U$. Each such edge $e \in M$ has $\deg_e(H) \geq (1-\epsilon) \cdot \beta$. This implies that $\deg_S(H) := \sum_{v \in S} \deg_v(H) \geq 4 \epsilon \cdot \mu(G) \cdot (1-\epsilon) \cdot \beta$. Next, observe that  each edge in $H$ that is incident on $S$ must share one of its endpoints with $\bar{A} \cup B$, and $\deg_v(H) \leq \beta$ for all nodes $v \in V$. Hence, we get $\mu(H) \cdot \beta = |\bar{A} \cup B| \cdot \beta \geq \deg_S(H) \geq 4\epsilon \cdot (1-\epsilon) \cdot \beta \cdot \mu(G)$, so that $\mu(H) \geq 4\epsilon \cdot (1-\epsilon) \cdot \mu(G) > \epsilon \cdot \mu(G)$. This leads to a contradiction.
\end{proof}

It now remains to extend Claim~\ref{cl:dichotomy:100} to general graphs. Towards this end, fix any general graph $G = (V, E)$ and an EDBS $H \subseteq E$ of $G$. Let $U \subseteq E \setminus H$ denote the collection of edges that are underfull w.r.t.~$H$. Let $M^* \subseteq E$ be a maximum matching in $G$. Following the approach taken by~\cite{AssadiB19}, we now construct a random bipartite subgraph $\hat{G} = (L \cup R, \hat{E})$ of $G$, as described below.

\medskip
\noindent {\bf Constructing the bipartite subgraph $\hat{G} = (L \cup R, \hat{E})$:} For each edge $e = (u,v) \in M^*$, sample a bit $\alpha_e \in \{0, 1\}$ uniformly at random: if $\alpha_e = 0$ then  assign $u$ to $R$ and $v$ to $L$,
else assign $u$ to $L$ and $v$ to $R$. Next, for each node $v \in V \setminus V(M^*)$, assign it to $R$ or $L$ uniformly at random. Finally, let $\hat{E} \subseteq E$ be the subset of edges in $G$ whose endpoints lie in different sides of the bipartition $(L, R)$. Note that $M^* \subseteq \hat{E}$, and hence $\mu(\hat{G}) = \mu(G) = |M^*|$. Define $\hat{H} := H \cap \hat{E}$.

\begin{claim}[\cite{AssadiB19}]
\label{cl:dichotomy}
  With non-zero probability,  the following conditions simultaneously hold.

\begin{itemize}
    \item (a) For every edge $e \in \hat{H}$, we have $\deg_{e}(\hat{H}) \leq (\beta/2) \cdot (1+4 \epsilon)$.
    \item (b) For every edge $e \in \hat{E} \setminus (\hat{H} \cup U)$, we have $deg_e(\hat{H}) \geq \beta/2 \cdot (1-5 \epsilon)$.
\end{itemize}
\end{claim}
For the rest of the proof, we condition on the event which ensures that both the two parts of Claim~\ref{cl:dichotomy} hold. Set $\hat{\epsilon} := 20 \epsilon$ and $\hat{\beta} := (\beta/2) \cdot (1+4\epsilon)$. Part~(a) of Claim~\ref{cl:dichotomy} implies that  $\hat{H}$ is an EDBS of $\hat{G}$, w.r.t.~$\hat{\beta}$ and $\hat{\epsilon}$. Let $\hat{U} := \left\{ e \in  \hat{E} \setminus \hat{H} : \deg_e(\hat{H}) < (1- \hat{\epsilon}) \cdot \hat{\beta} \right \}$ denote the collection of underfull edges in $\hat{G}$ w.r.t.~$\hat{H}$. Part~(b) of Claim~\ref{cl:dichotomy} implies that $\hat{U} \subseteq U$. Finally, recall that by definition, we have $\hat{H} \subseteq H$. Now, applying Claim~\ref{cl:dichotomy} on $(\hat{G}, \hat{H}, \hat{U})$, we get: 
\begin{equation}
    \label{eq:dichotomy:1}
\text{Either } \mu(\hat{G}) \leq (1/\hat{\epsilon}) \cdot \mu(\hat{H})  \text{ or } \mu(\hat{G}) \leq (1+\Theta(\hat{\epsilon})) \cdot \mu(\hat{H} \cup \hat{U}).
\end{equation}
Since $\mu(G) = \mu(\hat{G})$, $\hat{H} \subseteq H$, $\hat{U} \subseteq U$ and $\hat{\epsilon} = \Theta(\epsilon)$, the lemma  follows from~(\ref{eq:dichotomy:1}).

\bibliographystyle{alpha}
\bibliography{bibliography}

\appendix

\section{Proof of Theorem~\ref{thm:local-moves}}

\label{appendix:proof:theorem:edcs}

This argument is almost identical to the one present in \cite{AssadiB19}. Let $H \subset E$ be some arbitrary sub-graph. Define potential function $\Phi(H) = \sum_{v \in V} deg_H(v) \cdot (\beta - 1/2) - \sum_{e \in H} deg_H(e)$. Assume that $\epsilon \cdot \beta \geq 1$ and $\epsilon \cdot \beta$ is an integer for sake of simplicity. We will argue that if an edge $e^*$ in $E \setminus H$ which is underfull with respect to $H$ is added to $H$ or an edge $e^*$ of $H$ which is overfull with respect to $H$ is removed from $H$ then $\Phi(H)$ increases by at least one. Let $H^-$ and $H^+$ represent the state of $H$ before and after the edge update. In the former case: 

\begin{eqnarray}
\Phi(H^+) - \Phi(H^-) & = & \sum_{v \in V}(\beta - 1/2) \cdot (deg_{H^+}(v) - deg_{H^+}(v)) - \sum_{e \in H^+} deg_{H^+}(e) + \sum_{e \in H^-}deg_{H^-}(e) \nonumber \\
& = & 2 \cdot \beta - 1 - deg_{H^+}(e^*) - \sum_{e \in H^-} deg_{H^+}(e) - deg_{H^-}(e) \nonumber \\
& \geq & 2 \cdot \beta - 1 - (\beta \cdot (1-\epsilon) + 1) - (\beta \cdot (1-\epsilon) - 1) \nonumber \\
& \geq & -1 + 2 \cdot \beta \cdot \epsilon \nonumber \\
& \geq & 1 \nonumber
\end{eqnarray}

The first inequality just follows from the fact that $e^*$ had to be underfull before the insertion. Where as in the later case 

\begin{eqnarray}
\Phi(H^+) - \Phi(H^-) & = & - 2 \cdot \beta + 1 + deg_{H^+}(e^*) + \sum_{e \in H^-} deg_{H^+}(e) - deg_{H^-}(e) \nonumber \\
& \geq & - 2 \cdot \beta + 1 + (\beta + 1) + (\beta - 1) \nonumber \\
& \geq & 1 \nonumber
\end{eqnarray}

The first inequality follows simply from the fact that $e^*$ had to be overfull before the deletion. Hence, underfull edge insertions and overfull edge deletions consistently increase the potential function $\Phi(H)$. If $H$ is the empty-set $\Phi(H) = 0$. If there is no overfull edge in $H$ (with respect to $H$) then the maximum edge degree and hence the maximum vertex degree of $H$ is $\beta$. The fractional matching $f : H \Rightarrow 1/\beta$ is a feasible fractional matching and $|H|/\beta = size(f) \leq \mu(G) \cdot 3/2$, as the size of any fractional matching is at most $3/2$-the size of the maximum matching size in a graph. Furthermore, by definition $\Phi(H) = \sum_{v \in V}deg_H(v) \cdot (\beta-1/2) - \sum_{e \in H} deg_H(e) \leq \beta \cdot \sum_{v \in V}deg_H(v) = O(|H|\cdot \beta)$. The combination of the lower and upper bounds imply that $\Phi(H) \in O(\mu(G) \cdot \beta^2)$ and hence $H$ may undergo at most $O(\mu(G) \cdot \beta^2)$ underfull edge insertions or overfull edge deletions. 

Note that the only difference between the argument presented here and in \cite{AssadiB19} is that we express the upper bound on $\Phi(H)$ in terms of $\mu(G)$ instead of $n$ which turns out to be crucial in order to design our sub-linear approximate matching algorithm using adjacency list queries witouth additive slack.

\section{Missing proofs from Section~\ref{sec:implementation}}

\label{appendix:missingproofs:approx-matching}

\subsection{Proof of Lemma~\ref{lem:primitive:1}}

Observe that $X_i$ form i.i.d. variables, hence we can apply Chernoff's bound on $X$. By definition $\E[X] = \frac{S \cdot m}{n^2} = \frac{m \cdot \log(n) \cdot 100}{n \cdot \epsilon^3}$. Assume $m \geq \frac{\epsilon \cdot n}{2}$ hence $\E[X] \geq \frac{50 \cdot \log(n)}{\epsilon^2}$.

\begin{eqnarray}
\Pr(m \leq \hat{m} \leq m \cdot (1+ \epsilon) + 2 \cdot\epsilon \cdot n) & \geq & \Pr\left(m \in [\frac{X \cdot n^2}{(1+\epsilon) \cdot S}, (1 + \epsilon) \cdot \frac{X \cdot n^2}{S}] \right) \nonumber \\
 & = & \Pr\left(X \in [\frac{\E[X]}{1 + \epsilon}, \E[X] \cdot (1+\epsilon)]\right) \nonumber \\
& \geq & \Pr\left(|X - \E[X]| \leq \E[X] \cdot \frac{\epsilon}{2}\right) \label{eq:proof:lem:primitive:1:1} \\
& \geq & 1 - 2 \cdot \exp\left(-\frac{\E[X] \cdot (\epsilon/2)^2}{3}\right) \label{eq:proof:lem:primitive:1:2} \\
& \geq & 1 - 1/poly(n) \nonumber
\end{eqnarray}

Inequality~\ref{eq:proof:lem:primitive:1:1} holds as long as $\epsilon \leq 1/4$ and Inequality~\ref{eq:proof:lem:primitive:1:2} follows from Chernoff's bound. If $m < \frac{\epsilon \cdot n}{2}$ then $\E[X] \leq \frac{50 \cdot \log(n)}{\epsilon^2}$ and clearly $m \leq (1+\epsilon) \cdot \frac{X \cdot n^2}{S} + 2 \cdot \epsilon \cdot n = \hat{m}$. It remains to show that with high probability $\hat{m} \leq m \cdot (1+\epsilon) + 2 \cdot \epsilon \cdot n$, more specifically that with high probability $(1+\epsilon) \cdot \frac{X \cdot n^2}{S} \leq \epsilon \cdot n$ (and hence $\hat{m} \leq 2 \cdot \epsilon \cdot n$). 

\begin{eqnarray}
\Pr\left((1+\epsilon) \cdot \frac{X \cdot n^2}{S} \leq \epsilon \cdot n\right) & = & \Pr\left(X \geq \frac{100 \cdot \log(n)}{\epsilon^2 \cdot (1+\epsilon)}\right)\nonumber \\
& \leq & \Pr\left(X - \E[X] \geq \frac{50 \cdot \log(n)}{\epsilon^2} \cdot \frac{1-\epsilon}{1+\epsilon} \right)\nonumber \\
& \leq & \exp\left(-\frac{\frac{50 \cdot \log(n)}{\epsilon^2} \cdot (\frac{1-\epsilon}{1+ \epsilon})^2}{3}\right) \label{eq:proof:lem:primitive:1:3} \\
& \leq & 1/poly(\epsilon) \nonumber
\end{eqnarray}

Inequality~\ref{eq:proof:lem:primitive:1:3} follows from Chernoff's bound and from the fact that $\E[X] \leq \frac{50 \cdot \log(n)}{\epsilon^2}$.

\subsection{Proof of Lemma~\ref{lem:primitive:2}}

Let $X_e$ stand for the event where the algorithm samples edge $e$ and let $X_E$ stand for the event where the algorithm samples an edge from $E$. A simple application of Bayes's theorem shows that $\Pr\left(X_e|X_E\right) = \frac{\Pr(X_E | X_e) \cdot \Pr(X_e)}{\Pr(X_E)} = \frac{1 \cdot 1/n^2}{m/n^2} = 1/m$ as expected. The probability that any sample from $V \times V$ is not in $E$ is $(1-m/n^2)$. Hence, the probability that the algorithm doesn't sample an edge from $E$ in $\Omega\left(\frac{m \cdot \log(n)}{n^2} \right)$ samples is at most $(1-m/n^2)^{\Omega(\frac{m \cdot \log(n)}{n^2} )} \leq \exp(-\Omega(\log(n)) = 1/poly(n)$. 

\subsection{Proof of Theorem~\ref{th:main:adj-list:hybrid}}
\label{sec:th:main:adj-list:hybrid}

The proof of Theorem~\ref{th:main:adj-list:hybrid} is quite similar to the arguments developed in Section~\ref{sec:adj-list}. As usual, we essentially implement the schematic algorithm described in Figure~\ref{alg:sample}, under adjacency-list queries. The algorithm consists of the following steps.
 
\medskip
\noindent {\bf Step 1: Preprocessing.} This remains exactly the same as in Step 1 in Section~\ref{sec:adj-list}.

\medskip
\noindent {\bf Step 2: Obtaining a coarse approximation of $\mu(G)$.} We set $\Delta^* \leftarrow d$ (whereas Step 2 in Section~\ref{sec:adj-list} sets $\Delta^* \leftarrow \Delta$). Everything else remains the same as in Step 2 in Section~\ref{sec:adj-list}.

\medskip
\noindent {\bf Step 3: Obtaining the EDBS $H$.} This remains exactly the same as Step 3 in Section~\ref{sec:adj-list} (except that $\Delta^* = d$ instead of $\Delta$). Following the analysis in Section~\ref{sec:adj-list}, we infer that for this step the algorithm makes $\tilde{O}_{\epsilon}(m/d^{\gamma}) = \tilde{O}_{\epsilon}(n \cdot d^{1-\gamma})$ queries.

\medskip
At this point, we diverge from the algorithm in Section~\ref{sec:adj-list}, for reasons outlined in Section~\ref{sec:adj-list:hybrid}. Define the subset of nodes $V^* :=  \left\{ v \in V : \deg_v(E) \leq d/\epsilon \right\}$. Let $E^* := \{ (u, v) \in E : u, v \in V^*\}$ denote set of edges in $G$ induced by $V^*$. Next, define $G^* := (V^*, E^*)$, $H^* := H \cap E^*$ and $U^* := U \cap E^*$ (see Step~(13) in Figure~\ref{alg:sample}). Finally, define a subset of nodes $V^*_{small} \subseteq V^*$ such that:
\begin{equation}
    \label{eq:v:small:star}
    \left\{ v \in V^* : \deg_v(U^*) \leq \frac{(1-\epsilon) \cdot d^{\gamma}}{\epsilon} \right\} \subseteq V_{small}^* \subseteq \left\{ v \in V^* : \deg_v(U^*) \leq \frac{(1+\epsilon) \cdot d^{\gamma}}{\epsilon} \right\}.
\end{equation}
Finally, define $E^*_{small} := \left\{ (u, v) \in H^* \cup U^* : u, v \in V^*_{small}\right\}$, and let $G^*_{small} := (V^*_{small}, E^*_{small})$. The next two lemmas should respectively be interpreted as an analogues of Lemma~\ref{lem:approx} and Lemma~\ref{lem:maxdeg:schematic}.

\begin{lem}
\label{lem:hybrid:100} We have $\mu(G) \leq (3/2 + \epsilon) \cdot \mu(E^*_{small}) + \Theta(\epsilon n)$, whp.
\end{lem}

\begin{proof}
As $\deg_v(E) > d/\epsilon$ for all  $v \in V \setminus V^*$, we have
$\sum_{v \in V} \deg_v(E) = n \cdot d \geq |V\setminus V^*| \cdot (d/\epsilon)$, which implies that $|V \setminus V^*| \leq  \epsilon n$. Fix any maximum matching $M$ in $H \cup U$. Let $M^* \subseteq M$ be the set of edges in $M$ whose both endpoints are in $V^*$. Observe that $M^*$ is a valid matching in $H^* \cup U^*$. Since every edge in $M \setminus M^*$ shares at least one endpoint with $V \setminus V^*$, we have $|M \setminus M^*| \leq |V \setminus V^*|$. Thus, we get:
$\mu(H^* \cup U^*) \geq |M^*| = |M| - |M \setminus M^*| \geq |M| - |V \setminus V^*| \geq |M| - \epsilon n = \mu(H \cup U) -  \epsilon n$.
Rearranging the terms in the preceding inequality, we get:
\begin{equation}
    \label{eq:hybrid:1}
    \mu(H \cup U) \leq \mu(H^* \cup U^*) + \epsilon n.
\end{equation}
From~(\ref{eq:hybrid:1}) and  Theorem~\ref{thm:EDCS}, we infer that:
\begin{equation}
    \label{eq:hybrid:100}
  \mu(G) \leq (3/2 + \epsilon) \cdot \mu(H^* \cup U^*) + \Theta(\epsilon n).  
\end{equation}

It now remains to upper bound $\mu(H^* \cup U^*)$ by $\mu(E^*_{small})$.  Towards this end, recall that $\mu^* \leq n$ (see~(\ref{eq:parameters})),~$\Delta^* = d$ (see Step (2) above), and $U^* \subseteq U$. Hence, by Corollary~\ref{cor:deg:bound}, we have: $ |U^*| \leq n \cdot d^{\gamma}$, whp. Each node $v \in V^* \setminus V^*_{small}$ has $\deg_v(U^*) > (1-\epsilon) d^{\gamma}/\epsilon$. Accordingly,  we infer that
$|V^* \setminus V^*_{small}| \cdot (1-\epsilon) d^{\gamma}/\epsilon \leq |U^*| \leq  n  d^{\gamma}$, and hence $|V^* \setminus V^*_{small}| \leq  \epsilon (1-\epsilon)^{-1} n = \Theta(\epsilon n)$.

The edge-set $E^*_{small}$ is obtained by removing, from the set $H^* \cup U^*$, all the edges incident on the nodes in $V^* \setminus V^*_{small}$. Thus, applying an argument similar to the one used to derive~(\ref{eq:hybrid:1}), we get: 
\begin{equation}
    \label{eq:hybrid:500}
    \mu(E^*_{small}) \geq \mu(H^* \cup U^*) - |V^* \setminus V^*_{small}|\geq \mu(H^* \cup U^*) - \Theta(\epsilon n).
\end{equation}

The lemma follows from~(\ref{eq:hybrid:100}) and~(\ref{eq:hybrid:500}).
\end{proof}

\begin{lem}
\label{lem:maxdeg:hybrid} The graph $G^*_{small} := (V^*_{small}, E^*_{small})$ has maximum degree $\Delta_{G^*_{small}} = O_{\epsilon}\left( d^{\gamma}\right)$.
\end{lem}

\begin{proof}
Consider any node $v \in V^*_{small}$. By definition, we have $\deg_v(U^*) = O_{\epsilon}\left(d^{\gamma}\right)$. On the other hand, since $H$ is an EDBS of $G$, we have $\deg_e(H) \leq \beta$ for all edges $e \in H$, and hence $\deg_u(H) \leq \beta$ for all nodes $u \in V$. In particular, this means that $\deg_v(H) \leq \beta$, and hence $\deg_v(E^*_{small}) \leq  \deg_v(H) + \deg_v(U^*) \leq  O_{\epsilon}\left(\beta + d^{\gamma}\right) = O_{\epsilon}\left(d^{\gamma}\right)$. This implies the lemma.
\end{proof}

Recall that we already know the degree of each node in $G$ and the value of $d$ (see Step 1 above), as well as the contents of the set $H$ (see Step 3 above). Using this information, we explicitly identify the contents of the sets $V^*$ and $H^*$, without making any further adjacency-list query in $G$. This also implies that we have adjacency-matrix query access to the edges in $E^*$.

\medskip
\noindent {\bf Step 4: Identifying the set $V^*_{small}$.} This is analogous to Step 4 in Section~\ref{sec:adj-list}, with minor adjustments for the fact that the nodes in $V^*$ have maximum degree $d/\epsilon$ (as opposed to $\Delta$) in $G$. But, for completeness, we reproduce the appropriately modified arguments below.

For each node $v \in V^*$, we sample $K = (100/\epsilon^2) \cdot d^{1-\gamma} \log n$ edges in $G$ incident on $v$, uniformly and independently at random (with repetitions). We need to make one adjacency-list query in $G$ for each of these samples, which leads to a total of $\tilde{O}_{\epsilon}(d^{1-\gamma})$ adjacency-list queries per node, and hence $\tilde{O}_{\epsilon}(n \cdot d^{1-\gamma})$ adjacency-list queries to deal with all the nodes in $V^*$. For each $i \in [K]$, let $(v, u_i) \in E$ denote the $i^{th}$ sample for $v \in V$, and let $X_{i, v} \in \{0, 1\}$ be an indicator random variable that is set to $1$ iff $(v, u_i) \in U^*$. Note that once we have retrieved the sampled edge $(v, u_i) \in E$, it is easy for us to check whether it belongs to the set $U$ (by verifying whether $(v, u_i) \in E^* \setminus H^*$ and whether $\deg_v(H^*) + \deg_{u_i}(H^*) \leq (1-\epsilon) \beta$) without making any further adjacency-list queries in $G$, since we explicitly store all the edges in $H$ and have adjacency-matrix query access to the edges in $E^*$. Let  $X_v := \sum_{i=1}^K X_{i, v}$. We construct the set $V^*_{small}$ as follows: $V^*_{small} := \{ v \in V : X_v \leq (100/\epsilon^2) \cdot \log n\}$. We now show that this is consistent with the way $V^*_{small}$ is defined in~(\ref{eq:v:small:star}).

\begin{claim}
\label{cl:hybrid:degree}
The set $V_{small}$, as constructed above, whp satisfies the condition: 
$$\left\{ v \in V^* : \deg_v(U^*) \leq \frac{(1-\epsilon) \cdot d^{\gamma}}{\epsilon} \right\} \subseteq V^*_{small} \subseteq \left\{ v \in V^* : \deg_v(U^*) \leq \frac{(1+\epsilon) \cdot d^{\gamma}}{\epsilon} \right\}.$$
\end{claim}

\begin{proof}
Fix any node $v \in V^*$. The random variables $\{X_{i, v}\}$ are mutually independent, and we have $\mathbf{E}[X_v] = \sum_{i=1}^K \mathbf{E}[X_{i, v}] = \sum_{i=1}^K \Pr[X_{i, v} = 1] = K \cdot \deg_v(U^*)/\deg_v(E) = (100/\epsilon^2) \cdot d^{1-\gamma} \log n \cdot \deg_v(U^*)/\deg_v(E) \geq (100/\epsilon) \cdot d^{-\gamma} \log n \cdot \deg_v(U^*)$.\footnote{The last inequality holds because $\deg_v(E) \leq d/\epsilon$ for all nodes $v \in V^*$.} Thus, applying Chernoff bounds, we get:
\begin{obs}
\label{ob:hybrid:first}
If  $\deg_v(U^*) \leq \frac{(1-\epsilon) \cdot d^{\gamma}}{\epsilon}$, then  whp $v \in V_{small}$.  On the other hand, if  $\deg_v(U^*) > \frac{(1+\epsilon) \cdot d^{\gamma}}{\epsilon}$, then  whp $v \notin V_{small}$.
\end{obs}
The claim now follows from Observation~\ref{ob:hybrid:first} and a union bound over all the nodes $v \in V$.
\end{proof}

\medskip
To summarize, at the end of Step 4, we explicitly store the edge-set $H^*$ and  the node-set $V^*_{small}$, but {\em not} the edge-set $E^*_{small}$. However, given a pair of nodes $u, v \in V^*_{small}$, along with a guarantee (which we are in no position to verify) that $u$ and $v$ are connected via an edge in $E$, we can decide whether or not $(u, v) \in E^*_{small}$ based on the contents of the set $H^*$.

\medskip
\noindent {\bf Step 5: Computing a $(1+\epsilon)$-approximate estimate of $\mu(E^*_{small})$.} We now run the algorithm from Theorem~\ref{thm:Yoshida} on $G^*_{small} := (V^*_{small}, E^*_{small})$, and output the estimate $\alpha$ returned by this algorithm. The key challenge here is that Theorem~\ref{thm:Yoshida} requires adjacency-list query access to $G^*_{small}$, but we only have adjacency-list query access to $G$. We overcome this challenge as follows.

  Suppose that  this algorithm makes an adjacency-list query $Q$, asking for the $i^{th}$ edge incident on a given node $v \in V^*_{small}$ in $G^*_{small}$. To answer this  query $Q$, we make $\deg_v(E) \leq n$  adjacency-list queries in $G$ -- to collect all the edges incident on $v$ in $G$ -- and then identify the subset of these edges that belong to $G^*_{small}$ (based on the contents of  $H^*$ and $V^*_{small}$, which we explicitly store). Once we have identified this subset, we can trivially answer the original  query  $Q$.  Thus, we can implement each adjacency-list query in $G^*_{small}$ by making at most $n$ adjacency-list queries in $G$.

By Lemma~\ref{lem:maxdeg:hybrid} and Claim~\ref{cl:hybrid:degree}, whp the  graph $G^*_{small}$ has maximum-degree $\Delta_{G^*_{small}} = O_{\epsilon}(d^{\gamma})$. Accordingly, it follows that the algorithm from Theorem~\ref{thm:Yoshida}, while running on the graph $G^*_{small}$, makes $\tilde{O}_{\epsilon}\left(n \cdot (\Delta_{G^*_{small}})^{\Theta(1/\epsilon^2)}\right) = \tilde{O}_{\epsilon}\left( n \cdot d^{\Theta(\gamma/\epsilon^2)} \right)$ adjacency-list queries in $G$, whp.

\medskip
\noindent {\bf Approximation Guarantee:}
As discussed above, our algorithm returns an estimate $\alpha$ s.t.~$\alpha \leq \mu(E^*_{small}) \leq (1+\epsilon) \cdot \alpha$. Hence, Lemma~\ref{lem:hybrid:100} implies that:
\begin{equation}
\label{eq:hybrid:approx:main}
\alpha \leq \mu(E^*_{small}) \leq \mu(G) \leq  (3/2 + \Theta(\epsilon)) \cdot \alpha + \Theta(\epsilon n) = (3/2) \cdot \alpha + \Theta(\epsilon n).
\end{equation}

\medskip
\noindent {\bf Proof of Theorem~\ref{th:main:adj-list:hybrid}:}
The approximation guarantee follows from~(\ref{eq:hybrid:approx:main}). Next, recall that Steps 1, 2, 3, 4, 5 respectively makes  $\tilde{O}(n)$, $\tilde{O}_{\eps}(n)$, $\tilde{O}_{\eps}(n \cdot d^{1-\gamma})$, $\tilde{O}_{\eps}(n \cdot d^{1-\gamma})$ and $\tilde{O}_{\eps}\left(n \cdot d^{\Theta(\gamma/\epsilon^2)}\right)$  adjacency-list queries in $G$, whp. Hence, the total query complexity of the algorithm is at most:
\begin{eqnarray*}
\tilde{O}(n) + \tilde{O}_{\eps}(n) + \tilde{O}_{\eps}(n \cdot d^{1-\gamma}) +\tilde{O}_{\eps}(n \cdot d^{1-\gamma}) + \tilde{O}_{\eps}\left(n \cdot  d^{\Theta(\gamma/\epsilon^2)}\right)    = \tilde{O}_{\epsilon}\left( n \cdot \left(1 + d^{1-\Theta(\epsilon^2)}\right) \right).
\end{eqnarray*}
This holds because $\gamma := c \cdot \epsilon^2$ for a sufficiently small constant $c > 0$ (see Step 2 above). \qed

\section{Efficient implementation of the framework}

\label{sec:appendix:runtimeyoshida}

In this section we will show that the algorithms presented in this paper can be implemented in running time roughly proportional to their query complexity. We will first discuss the running time of implementing the schematic algorithm. Based on the established query complexity bounds presented in the paper it is sufficient to argue that the total computational work accompanied with each edge sample is relatively low.

The first part of the schematic algorithm finds the edges of $H$. For each sampled edge $e \in E$ to determine weather the given edge should be added to $H$ the algorithm simply needs to query $deg_H(e)$. This can be done in constant time as $H$ is stored explicitly. If $e$ is added to $H$ the algorithm needs to make sure that overfull edges get removed from $H$. The addition of $e$ may only increase the edge degree of edges neighbouring $e$. Hence it is enough to check the $O(\beta) = O(1/poly(\epsilon))$ neighbouring edges of $e$ for overfull-ness after each edge insertion.

The second part of the schematic algorithm needs to find $V_{small}$. This is achieved through the random sampling of edges incident on all vertices of $V$. Throughout this sampling process the algorithm simply needs to check the membership of edges in $e$ which can be done in constant time. The total amount of computational work completed corresponds to simply updating counters after each sample hence the time complexity here will correspond to the sample complexity. One small vertices are determined for any edge $e$ present in $E$ the algorithm can efficiently decide membership in $H \cup U$ as this only requires to check the membership of the endpoints of $e$ in $V_{small}$ and the degree of $e$ in $H$.

It remains to prove Theorem~\ref{thm:Yoshida}. In \cite{YoshidaYI12} the authors present algorithms with bounds on their edge query complexity. Through following their analysis similar bounds on running-time also follow, however that would require significant technical effort. For sake of succinctness in this section we show that the running time bounds also follow from their query complexity bounds alone. Everything present in this section can be considered as a corollary of \cite{YoshidaYI12} and \cite{Behnezhad21}.

The authors in \cite{YoshidaYI12} first introduce an algorithm which efficiently computes the membership of an edge chosen uniformly at random in some maximal independent set of graph. Afterwards, the authors show how the membership of a uniformly random vertex in some fixed $(1+\epsilon)$-approximate matching may be calculated using the MIS algorithm as a sub-routine. 

We will need to slightly transform their algorithm to be able to query the matched status of a random vertex instead of edge under a $(1+\epsilon)$-approximate matching. Using the later algorithm we may obtain a $(1+\epsilon)$-approximation to the maximum matching size using techniques presented in \cite{Behnezhad21}. Note that the aim of \cite{YoshidaYI12} was to design an approximate maximum size estimate in expected query complexity independent of $n$. To achieve this their algorithm randomly samples the matched status of some $O_\epsilon(\Delta^2)$ edges of the graph under some fixed $(1+\epsilon)$-approximate matching and creates an estimate based on the ratio of matched edges within the sample. This approach yields a $(1+\epsilon,\epsilon \cdot n)$-approximate matching size estimate. However, similarly as in \cite{Behnezhad21} we may instead sample some $O_\epsilon(\Delta \cdot \log(n))$ uniform vertices to obtain an approximation ratio of $(1+\epsilon)$.

\begin{lem}

\label{lm:appendix:yoshida:MIS}

There is a randomized algorithm $\mathcal{A}_{MIS}$ that given graph $G$ with maximum degree $\Delta$ calculates the membership of a uniformly sampled random vertex of $G$ under some fixed maximal independent set of $G$ in expected running time of $O(\Delta^2)$ using adjacency list queries on the edges of $G$.

\end{lem}

\begin{lem}

\label{lm:appendix:yoshida:matching}

There is a randomized algorithm $\mathcal{A}_{1+\epsilon}$ that given graph $G$ with maximum degree $\Delta$ calculates the membership of a uniformly sampled edge of $G$ under some fixed $(1+\epsilon)$-approximate matching of $G$ in expected running time of $O(\Delta^{1/\epsilon^2})$ using adjacency list queries on the edges of $G$.

\end{lem}

\begin{cor}

\label{cor:appendix:yoshida:matching:vertex}

There is a randomized algorithm $\mathcal{A}_{1+\epsilon}$ that given graph $G$ with maximum degree $\Delta$ calculates the membership of a uniformly sampled vertex of $G$ under some fixed $(1+\epsilon)$-approximate matching of $G$ in expected running time of $O(\Delta^{1/\epsilon^2})$ using adjacency list queries on the edges of $G$.

\end{cor}

\subsection{Proof of Lemma~\ref{lm:appendix:yoshida:MIS}}

\label{sec:appendix:proof:MIS}

The goal of this subsection is to argue that the running time of the algorithm for MIS membership in \cite{YoshidaYI12} is the same as it's query complexity. While the statement follows as an elementary corollary of \cite{YoshidaYI12} the authors did not specify their results in terms of running time but edge query complexity.

A greedy maximal independent set is defined of a graph $G = (V,E)$ is defined by some ordering $\pi$ of the vertices of $V$. We may sequentially find a greedy MIS as follows: initialize the MIS $S$ as the empty-set and iterate through the vertices of $V$ with respect to $\pi$. For any $v$ if no neighbour $v'$ of $v$ was added to $S$ before reaching $v$ add $v$ to $S$.

Define oracle $\mathcal{O}_\pi$ to take some vertex $v$ and return it's matched status under the greedy MIS defined by $\pi$. Oracle $\mathcal{O}_\pi$ for vertex $v$ is implemented as follows: iterate through the neighbours of $v$ (in some arbitrary fixed order). If any neighbour $v'$ of $v$ if $\pi(v') < \pi(V)$ and $v'$ is in the MIS (which we query through $\mathcal{O}_{\pi}$) we return that $v$ is not in the MIS. If we find no such neighbour of $v$ we add it to the MIS. Observe that aside from the recursive calls to $\mathcal{O}_{\pi}$ the oracle may be implemented in $O(\Delta)$ time (and adjacency list queries) for any vertex $v \in V$ given $\pi$. Furthermore, when accessing $v$ to make the appropriate $\mathcal{O}_\pi$ calls on it's neighbours it is sufficient to evaluate $\pi$ on the neighbours of $v$. This can be done efficiently through drawing $O(\log(n))$ long random bit series and assigning them to vertices of $V$ as $\pi(v)$ locally. 

\cite{YoshidaYI12} has elegantly shown that given $v$ is chosen uniformly at random a call to $\mathcal{O}_\pi(v)$ in expectation (with respect to the randomness of sampling $v$ and $\pi$) will trigger $O(\Delta)$ calls to $\mathcal{O}_\pi(v)$ on different vertices (Theorem 2.1). Hence, in expectation the recursive calls to $\mathcal{O}_{\pi}$ may be simulated in $O(\Delta^2)$ time when accessing a random a vertex.

\subsection{Proof of Lemma~\ref{lm:appendix:yoshida:matching}}

In this subsection we will explain for the sake of completeness how \cite{YoshidaYI12} uses the algorithm of Lemma~\ref{lm:appendix:yoshida:MIS} to implement an oracle which efficiently queries weather a random vertex is matched under some fixed $(1+\epsilon)$-approximate matching and show that the query complexity of their algorithm is roughly the same as it's running time.

We will now first describe the $(1+\epsilon)$-approximate matching \cite{YoshidaYI12}-s algorithm queries locally. An augmenting path with respect to some matching $M$ is a path of the graph consisting of edges alternating from $E\setminus M$ and $M$, starting and ending with an edge of $E \setminus M$. We will say that an augmenting path with respect to $M$ is eliminated if its edges are used to increase the size of $M$ by one through flipping their membership in $M$. 

Let $M_0$ be the empty-set. Generate $M_{i+1}$ from $M_i$ through eliminating a maximal set of vertex disjoint augmenting paths (in the first round this is simply a maximal matching) with respect to $M_i$ of length $(2i-1)$. It is well known by \cite{hopcroft} that the matching $M_i$ generated this way is $\frac{i}{i+1}$-approximate. The goal is to efficiently implement an oracle $\mathcal{O}_k$ which determines the matched status of some edge $e$ in $M_k$ where $k = 1/\epsilon$. In order to achieve this goal \cite{YoshidaYI12} describes a series of oracles. Let $P_i$ stand for the set of $(2i-1)$ long paths of $G$. Let $V_i \subseteq P_i$ stand for the subset of paths of $P_i$ which are augmenting paths with respect to $M_{i-1}$.  Let $G_i = (V_i, E_i)$ where there is an edge between pairs of vertices $(v_i, v_i') \in V_i \times V_i$ if the corresponding augmenting paths in $G$ share a vertex. Let $A_i$ stand for an MIS of $G_i$ generated by the algorithm of Lemma~\ref{lm:appendix:yoshida:MIS}. Using this construction we may define $M_0,\cdots,M_k$ as a function of $G,\pi_0, \cdots, \pi_k$ where $\pi_i$ stands for a permutation over the augmenting path vertices $V_i$. Let $\Delta_i$ stand for the maximum degree of $G_i$. By definition $\Delta_i \leq \Delta^{2i}$

The authors define 3 oracles (in a slightly different construction):

\begin{itemize}

    \item $\mathcal{O}_i : E \rightarrow $ True/False queries the membership of an edge of $E$ under $M_i$
    
    \item $\mathcal{H}_i : V_i \rightarrow $ True/False queries the membership of an augmenting path in $V_i$ under the MIS $A_i$
    
    \item $\mathcal{V}_i : P_i \rightarrow $ True/False queries the membership of path $p \in P_i$ in $V_i$
    
\end{itemize}

Queries are answered inductively. $P_1 = V_1 = E$ initially. Hence a query to $\mathcal{O}_1$ can be answered through running the algorithm of Lemma~\ref{lm:appendix:yoshida:MIS} on graph $G_1$, which is the line graph of $G$. Note, that we may assume that every edge is contained in some $(2i-1)$ long augmenting path. If this is not the case we may completely construct the component containing the edge in $\Delta^{O(i)}$ time and calculate membership in $M_i$ restricted to it. A query to $\mathcal{O}_i(e)$ is answered as follows:

\begin{itemize}

    \item The algorithm first finds all $(2i-1)$ long paths containing $e$ at some oddth position, call these paths $P^{e}_i$

    \item Afterwards the algorithm selects the paths of $P^{e}_i$ which are augmenting with respect to $M_{i-1}$, call these $V^e_i$, using $\mathcal{V}_i$

    \item Finally the algorithm determines if any of $V^e_i$ are in $A_i$ using $\mathcal{H}_i$

    \item Iff either A) a path of $V^e_i$ is in $A_i$ or B) $e \in M_{i-1}$ (tested by a single call to $\mathcal{O}_{i-1}$) $\mathcal{O}_i$ returns True.
    
\end{itemize}

A query to $\mathcal{H}_i(v_i)$ is executed through running the algorithm of Lemma~\ref{lm:appendix:yoshida:MIS} on $G_i$. This requires the evaluation of $O(\Delta_i^2)$ adjacency list queries on $G_i$. List query of $(v_i,j)$ requesting the $j$-th neighbour of $v_i$ in $E_i$ is executed as follows:

\begin{itemize}

    \item The algorithm first enumerates all augmenting $(2i-1)$ long paths in $P_i$ which share a vertex with $v_i$

    \item The algorithm then calls $\mathcal{V}_i$ on each of these paths to determine if they are augmenting paths with respect to $M_{i-1}$ and returns the $j$-th amongst them (according to some arbitrary but fixed ordering of $V_i$)
    
\end{itemize}

A query to $\mathcal{V}_i(p_i)$ is executed through querying all edges $e \in p_i$ with $\mathcal{O}_{i-1}$. \cite{YoshidaYI12} has shown that assuming we query a uniformly random edge of $E$ with $\mathcal{O}_k$ the previously described recursive process will subject $G$ to a total of $\Delta^{O(k^2)} = \Delta^{O(1/\epsilon^2)}$ adjacency list queries in expectation (Theorem 3.7). It remains to argue that the running time of executing the recursive process is roughly proportional to its query complexity. Note that the bound presented by \cite{YoshidaYI12} is pessimistic in a sense that it does not rely on caching results of earlier queries. Their bound assumes that the oracles may query the same edge of $G$ repeatedly without storing the previous information.

First note that these queries can be executed without the explicit knowledge of permutations $\pi_0, \cdots \pi_k$. A permutation a set $S$ can be locally generated through selecting random bit strings of length $O(\log(|S|))$ to represent elements of $S$. As $|V_i| \leq O(n^{2i})$ it is sufficient to draw $O(\log(n)/\epsilon)$ long bit strings in order to locally query the permutations $\pi_i$.

Following the analysis of \cite{YoshidaYI12} one can confirm that their bounds extend to the expectation of the running time. However, for sake of succinctness we may bound the running time of the recursive algorithm based on the authors bound on query complexity in a straightforward manner. 

We will first argue that the running time of a call to $\mathcal{O}_i$ or $\mathcal{H}_i$ (not accounting for the cost of subsequent oracle calls) is at most $O(k) \cdot \Delta^{O(k)}$ larger then their query complexity, even if we assume they query a single edge. Both methods need to enumerate sets of $(2i-1)$ long paths in $G$ (and make a constant number of further oracle calls on them). $\mathcal{O}_i$ needs to enumerate all such paths containing a specific edge while $\mathcal{H}_i$ needs to enumerate all such paths sharing a vertex with a $(2i-1)$-long path. These processes can be completed in $O(i \cdot \Delta^{2i-2})$ and $O(i \cdot \Delta^{4i-2})$ time through exhaustive breath first searches in a straightforward manner. Hence, their running time is at most $O(k)\cdot \Delta(O(k))$ larger then their query complexity. 

A call to $\mathcal{V}_i$ requires the algorithm to make $(2i-1)$ further oracle calls to $\mathcal{O}_{i-1}$, however it does not require a list-query of graph $G$. A call of $\mathcal{V}_i$ makes calls to $\mathcal{O}_{i-1}$. We may charge the relatively small cost of executing $\mathcal{V}_i$ to the subsequent $(2i-1)$ $\mathcal{O}_{i-1}$ calls they make. As in the chain of recursion $\mathcal{V}$-calls are accessing a lower layer oracle $\mathcal{O}_{i-1}$ this charges may just add up to the depth of the recursion, $O(k)$ (we may assume that any $\mathcal{O}_1$ call requires at least one edge query of $G$). Overall, the algorithm may complete at most $O(k) \cdot \Delta^{O(k)}$ more computational work then edge queries to the underlying graph hence its running time is bounded by $O(k \cdot \Delta^{k^2}) = O(\Delta^{1/\epsilon^2}/\epsilon)$.

\paragraph{Proof of Corollary~\ref{cor:appendix:yoshida:matching:vertex}:} In order to query the matched status of a vertex with the prior mentioned framework we randomly sample a vertex $v$ and query the matched status of all edges incident on $v$. We will argue that the expected running time of such vertex queries can be at most $O(\Delta)$ larger than of edges. Consider the probability that a specific edge is queried with the initial query $\mathcal{O}_k$ under the run of the two separate query processes. In the case of the algorithm of \cite{YoshidaYI12} it is $1/m$ as the sampling is uniform over the edges. In the case of vertex queries the same probability is $2/n = O(\Delta/m)$ corresponding to the probability that one of its endpoints is sampled. Hence, due to the linearity of expectations the expected cost of querying all edges incident on a random vertex with $\mathcal{O}_k$ should only be an $O(\Delta)$-factor more expensive then querying a random edge.

\subsection{Proof of Theorem~\ref{thm:Yoshida}}

\label{sec:appendix:proof:matching:estimate}

This section can be derived as a corollary of \cite{YoshidaYI12} by using techniques present in \cite{Behnezhad21}. First observe that $\mu(G) \geq \frac{n}{2 \cdot \Delta}$. This can be concluded from Vizings theorem or by the fact that we can define a valid fractional matching $f : E \rightarrow 1/\Delta$ on the edges of $G$ such that $size(f) \geq n/\Delta$ and it is folklore that for all valid fractional matchings $size(f) \leq \mu(G) \cdot 3/2$. 

The algorithms samples $T = \frac{\Delta \cdot \log(n) \cdot 10^5}{\epsilon^2}$ vertices of $V$ uniformly at random. The algorithm queries the membership of all sampled vertices in a $(1+\epsilon/8)$-approximate matching using the algorithm of Corollary~\ref{cor:appendix:yoshida:matching:vertex}. Let $X_i$ stand for the indicator variable of the event where the $i$-th sampled vertex belongs to the matching and $X = \sum X_i$. Note, that as $\mu(G) \geq n/(2 \cdot \Delta)$ we must have that $\E[X] = T \cdot n/(2\cdot \mu(G))\geq \frac{\log(n) \cdot 10^5}{\epsilon^2}$. The algorithm returns $\tilde{\mu} = X \cdot n \cdot (1-\epsilon/2)/(2 \cdot T)$ as the matching size estimate. By the guarantees of Corollary~\ref{cor:appendix:yoshida:matching:vertex} the algorithm runs in $\tilde{O}(\Delta^{1/\epsilon^2})$ time (and query complexity).

\begin{claim}

$\tilde{\mu} \in [\mu(G) / (1+\epsilon), \mu(G)]$ with high probability.

\end{claim}

\begin{proof}

We will first show that $X \in [\E[X] \cdot (1 -\epsilon/8), \E[X] \cdot (1+\epsilon/8)]$ with high probability. Note that $X_i$ are i.i.d. variables.

$$P(|X - \E[X]| \geq \E[X] \cdot \epsilon/8) \leq 2 \cdot \exp\left(\frac{-\E[X] \cdot \epsilon^2}{3} \right) \leq \exp(-10 \cdot \log(n)) \leq 1/poly(n)$$

The inequalities follow from a simple application of Chernoff's bound and the fact that $\E[X] \geq \frac{\log(n) \cdot 10^5}{\epsilon^2}$. $\E[X] \cdot n/(2 \cdot T) \in [\mu(G) \cdot (1-\epsilon/8), \mu(G)]$ by the assumption that the algorithm querying membership in a $(1+\epsilon/8)$-approximate matching. Hence, with high probability

$$\tilde{\mu} = \frac{X \cdot n \cdot (1-\epsilon/2)}{2 \cdot T} \leq \frac{\E[X] \cdot n\cdot (1-\epsilon/2)\cdot(1+\epsilon/8)}{2 \cdot T} \leq  \mu(G)(1-\epsilon/2) \cdot (1+\epsilon/8)\leq \mu(G)$$

$$\tilde{\mu} = \frac{X \cdot n \cdot (1-\epsilon/2)}{2 \cdot T} \geq \frac{\E[X] \cdot n \cdot  (1-\epsilon/2)\cdot(1-\epsilon/8)}{2 \cdot T} \geq \mu(G) \cdot (1-\epsilon/2)\cdot(1-\epsilon/8)^2 \geq \mu(G) / (1+\epsilon)$$

\end{proof}

Note that by the guarantees of Lemma~\ref{lm:appendix:yoshida:matching} the bound on the running time of the algorithm described above only holds in expectation (however it's output is correct with high probability). While incurring at most an extra $O(\log(n))$ factor this may be improved to high-probability using standard methods. In particular, any algorithm with an expected running time of $\E[\mathcal{A}]$ will compile in time $O(\E[\mathcal{A}])$ with probability $1/2$ by Markov's lemma. Hence, if the algorithm is run $\Theta(\log(n))$ times in parallel with probability $1 - 1/poly(n)$ one of it's copies compiles within $O(\E[\mathcal{A}])$ time. As the algorithms output was correct with probability $1-1/poly(n)$ all parallel copies should also return a correct solution with probability $1/poly(n)$.

\section{Sub-linear TSP}

\label{appendix:tsp}

In the metric traveling salesman problem we are given an $n \times n$ distance matrix $D$ specifying the pairs wise distance of $n$ locations and our goal is to devise a tour of minimum length visiting all locations. Similarly to the matching problem it is an interesting question how close the cost of the optimal solution can be approximated via a sub-linear number ($o(n^2)$) of queries to $D$. \cite{CzumajS09MST} has shown that the cost of a minimum size spanning tree of a weighted graph may be estimated within a $(1+\epsilon)$-factor using $\tilde{O}(n)$ queries. As the TSP cost is upper bounded by the double of the MST cost this approach yields an efficient $(2+\epsilon)$-approximate solution to the problem.

For specific metrics better than $2$-approximate solutions using sub-linear time have appeared in literature (\cite{MomkeS1612MST1,Mucha14MST2,SeboV14MST3}) (but not for general metrics which remains an interesting open problem). In particular for the case of graphic TSP where distances in $D$ represent vertex pair distances in an underlying unweighted graph \cite{chen2020sublinear} has shown that there is a $\tilde{O}(n)$ time algorithm which approximates the cost within a $2-\delta$ ratio for some absolute constant $\delta$. In the same paper the authors show that no algorithm can approximate the cost of graphic TSP within an approximation ratio close than $(1+\epsilon_0)$ for some absolute constant $\epsilon_0$ with $o(n^2)$ distance queries. Furthermore, the authors show that with query complexity super-linear in $n$ a better approximation ratio $27/14$ can be obtained in $\tilde{O}(n^{3/2})$ time. Note that this result relies on an approximate maximum matching algorithm which runs in $\tilde{O}(n^{3/2})$ which was recently improved to $\tilde{O}(n)$ time by \cite{behnezhad2022time}. 

There are sub-linear results for the case of $(1,2)$-TSP (where distances in $D$ are in $\{1,2\}$) (\cite{chen2020sublinear} $1.625$-approximation, \cite{AdamaszekMP1812MST} $1.75$-approximation) and when it is either known that the metric contains a spanning tree supported on weight-1 edges or the algorithm is given access to a minimum spanning tree of the graph (\cite{MSTgeneral} sub-2 approximation).

In \cite{chen2020sublinear} the authors show a close connection between the approximate matching and TSP problems. They show that any algorithm which can approximate the cost of graphic or $(1,2)$-TSP within a $(1+\epsilon)$-factor may be used to approximate the size of a maximum matching of a bipartite graph with $\epsilon \cdot n$ additive slack.

Substituting Theorem~\ref{thm:main} into the framework of \cite{chen2020sublinear} we achieve a new approximation ratio achievable for the TSP problem in sub-linear time: we may obtain a $40/21$-approximate solution in $\tilde{O}(n^{2-\theta{\epsilon^2}})$ time and distance queries. This suggest that there might indeed be an interesting regime of different approximation ratio/running time payoffs for TSP for approximation ratios in $(1+\epsilon_0,2)$ and running times $(n,n^2)$. 

\begin{thm}[Theorem 7 of \cite{chen2020sublinear}]

\label{thm:tsp-yuteal} For any $\delta$ and $c_0 \geq 1$. Given a graph G with maximum matching size $\alpha \cdot n$, suppose there is an algorithm that uses pair queries (adjacency matrix queries), runs in $t$ time, and with probability at least 2/3, outputs an estimate of the maximum matching size $\hat{\alpha} \cdot n$ such that $\hat{\alpha} \leq \alpha \leq c_0 \cdot \hat{\alpha} + \delta$. Then there is an algorithm that approximates the cost of graphic TSP of G to within a factor of $2-\frac{1}{7 \cdot c_0} + \delta$, using distance queries, in $t + O(\frac{n}{\delta^2})$ time with probability at least $3/5$.

\end{thm}

The breakthrough paper \cite{behnezhad2022time} has shown an $\tilde{O}(n)$ running time algorithm for finding a $(2+\epsilon)$-approximate matching size estimate. Substituting this algorithm into into Theorem~\ref{thm:tsp-yuteal} of \cite{chen2020sublinear} the author of \cite{behnezhad2022time} obtains a $27/14$-approximation to the graphic TSP-problem in $\tilde{O}(n)$ time. If we instead substitute the algorithm of Theorem~\ref{thm:main} we obtain the following result:

\begin{thm}

\label{thm:tsp}

There is an $\tilde{O}(n^{2-\Omega(\epsilon^2)})$ running time randomized algorithm which estimates the cost of graphic TSP within a factor of $40/21+\epsilon$.

\end{thm}

\section{Proof of Fact~\ref{fact:main}}
\label{appendix:fact}

\begin{wrapper}
\begin{fact}\label{fact:main}

(Formal:) There is no algorithm which for a given graph and some constant $\epsilon>0$ returns a $1/\epsilon$-approximate (respectively $(1/\epsilon, \epsilon \cdot n)$ approximate) matching in $\tilde{O}_{\epsilon}(n^{2-\delta})$ time for some $\delta>0$ using adjacency list query (respectively adjacency matrix query) access to the graph.

\end{fact}
\end{wrapper}

\label{appendix:fact}

\paragraph{Adjacency Matrix Queries:} Consider a bipartite graph on vertex sets $|L|=|R|=n$ where any potential edge $(u,v) | u \in L, v \in R$ exists with some probability $p = \frac{k \cdot \log(n)}{\epsilon^2}/n$ independently from each other for some large enough constant $k$. Call the edge set of this graph $E_p$. Using chernoff's bound we can argue that if $k$ is selected high enough then with high probability the degree of each vertex will fall into $[\frac{(1-\epsilon) \cdot \log(n) \cdot k}{\epsilon^2}, \frac{(1+\epsilon) \cdot \log(n) \cdot k}{\epsilon^2}]$ in $E_p$. Consider fractional matching $E_p \rightarrow n \cdot p/(1+\epsilon)$. As with high probability vertex degrees are in $[(1-\epsilon) \cdot p \cdot n, (1+\epsilon) \cdot p \cdot n]$ $w$ is a valid fractional matching and $size(w) \geq \frac{n \cdot (1-\epsilon)}{1+\epsilon} \geq size(w) \cdot n \cdot (1-4 \cdot \epsilon)$. As $w$ is a valid fractional matching and $G$ is bipartite $\mu(G) \geq size(w)$.

Any $(1/\epsilon, \epsilon \cdot (2n))$-approximate matching of $G$ must have at least $(\mu(G) - \epsilon \cdot (2n)) \cdot \epsilon \geq \epsilon \cdot n \cdot (1-6 \cdot \epsilon) \geq n \cdot \epsilon / 2$ edges (for $\epsilon < 1/12$). However, as each edge exist in the graph with probability $p$ independently from each other, regardless which edges are queried by the algorithm it needs to query (in expectation and by Chernoff's bound with high probability) at least $\Omega(\frac{n \cdot \epsilon}{p}) = \Omega_\epsilon(\frac{n^2}{\log(n)})$ edges in order to find a large enough matching.

\paragraph{Adjacency List Queries:} Construct graph $G = (L \cup R \cup U, E)$ as follows: $|L| = |R| = n$ and there is a perfect matching (and no other edge) between $L$ and $R$, $|U| = n \cdot \epsilon/2$ and there is a complete matching between $L \cup R$ and $U$. Assume that the neighbour list of every vertex is ordered uniformly at random. By definition $\mu(G) = n$. An algorithm may easily match the vertices of $U$ to some subset of vertices in $L \cup R$. However, in order to match $\epsilon \cdot n$ pairs of vertices the algorithm must find at least $n \cdot \epsilon/2$ edges of the perfect matching between $R$ and $L$. Any adjacency list query on the edges incident of a vertex in $R$ or $L$ returns a neighbour in $U$ with probability $(n\cdot\epsilon/2)/(n\cdot \epsilon/2 + 1)$ and returns a vertex of $R \cup L$ with probability $1/(n \cdot \epsilon/2 +1)$. Hence, in expectation the algorithm must make at least $\Omega(n \cdot \epsilon)$ list queries to find a single edge of the perfect matching. Therefore, to construct a matching of size $\epsilon \cdot n$ the algorithm must make at least $\Omega_\epsilon(n^2)$ list queries to edges incident on the vertices of $U \cup L$.

\paragraph{Note on the definition of adjacency list queries:} One can define adjacency list queries in two separate ways. In both settings query of the format $(v,i)$ returns the $i$-th neighbour incident on vertex $v$. However, the definition $i$-th neighbour can either follow from a pre-defined ordering of vertices over the whole graph or this ordering may be different for all vertices. Papers in literature are often ambiguous about this detail. In the former, easier setting the proof in the paragraph above breaks down (for that setting their is a binary search based algorithm which works on the above example graph). However, informally speaking if instead of $U$ being completely connected to $R \cup L$ we would define the graph such that any edge $(u,v) | u \in U, v \in R \cup L$ exists with probability $1/2$ the algorithm would still not have any better way to find an edge of the perfect matching then querying $\Omega(\epsilon \cdot n)$ edges of some vertex in $R \cup L$ and we would reach the same conclusion.

\end{document}